\documentclass{article}

\usepackage[a4paper, total={6in, 8in}]{geometry}

\usepackage[utf8]{inputenc} 
\usepackage[T1]{fontenc}    
\usepackage{hyperref}       
\usepackage{url}            
\usepackage{booktabs}       
\usepackage{amsfonts}       
\usepackage{nicefrac}       
\usepackage{microtype}      
\usepackage{lipsum}
\usepackage{array}
\newcolumntype{P}[1]{>{\centering\arraybackslash}p{#1}}

\usepackage{graphicx}
\usepackage{amsmath}
\usepackage{amssymb}
\usepackage{float}
\usepackage{subfig}

\usepackage[linesnumbered,ruled,vlined]{algorithm2e}
\SetKwInput{KwInit}{Input}

\title{Machine-learning energy-preserving nonlocal closures for turbulent fluid flows and inertial tracers}

\author{Alexis-Tzianni G. Charalampopoulos, Themistoklis P. Sapsis\footnote{Corresponding author: sapsis@mit.edu, Tel (617) 324-7508, Fax: (617) 253-8689}\\
Department of Mechanical Engineering, Massachusetts Institute of Technology}

\begin{document}

\maketitle

\begin{abstract}
We formulate a data-driven, physics-constrained closure method for coarse-scale numerical simulations of turbulent fluid flows. Our approach involves a closure scheme that is non-local both in space and time, i.e. the closure terms are parametrized in terms of the spatial neighborhood of the resolved quantities but also their history.
The data-driven scheme is complemented with a physical constrain expressing the energy conservation property of the nonlinear advection terms. We show that the adoption of this physical constrain not only increases the accuracy of the closure scheme but also improves the stability properties of the formulated coarse-scale model. 
We demonstrate the presented scheme in fluid flows consisting of an incompressible two-dimensional turbulent jet. Specifically, we first develop one-dimensional coarse-scale models describing the spatial profile of the jet. We then proceed to the computation of turbulent closures appropriate for two-dimensional coarse-scale models.  Training data are obtained through high-fidelity direct numerical simulations (DNS). We also showcase how the developed scheme captures the coarse-scale features of the concentration fields associated with inertial tracers, such as bubbles and particles, carried by the flow but not following the flow. We thoroughly examine the generalizability properties of the trained closure models for different Reynolds numbers, as well as, radically different jet profiles from the ones used in the training phase. We also examine the robustness of the derived closures with respect to the grid size. Overall the adoption of the constraint results in an average improvement of $26\%$ for one-dimensional closures and $29\%$ for two-dimensional closures, being notably larger for flows that were not used for training. 
\end{abstract}


 \section{Introduction}
 Turbulent fluid flows in nature and engineering are characterized by a wide range of spatial and temporal scales with nonlinear  interactions making their reduced order modeling a challenging task. Over the last decades several ideas have emerged that successfully model turbulent fluid flows, such as Large Eddy Simulations \cite{Moin1991, Sagaut2002}. However, these methods still require very high resolution in order to satisfactory model the large scale dynamics, as well as features associated with those. This is an important computational obstacle especially for applications involving uncertainty quantification, optimization, and risk analysis where there is a need for a large number of accurate simulations.

The recent machine-learning advances has sparked a new interest in  utilizing deep neural networks to develop reduced order models for turbulent flows. The machine-learning closures abandon the path of a closed-form expression for the closure terms into utilizing experimental or costly high-fidelity computations to train a neural network and predict the nonlinear energy transfers between resolved and unresolved scales. One of the first such effort of deep neural networks to turbulent flows appeared in \cite{milano2002neural}, where nonlinear autoencoders were utilized to reconstruct the near wall field in a turbulent flow. Since then there has been a plethora of efforts focusing on machine-learning closures using different data-driven schemes, such as artificial neural networks in fluid flows \cite{gamahara2017searching, singh2017machine, maulik2019subgrid, Yang2019} and multiphase flows \cite{ma2015using, ma2016using}, random forest regressions \cite{wu2018physics}, spatially nonlocal schemes such as convolutional neural networks \cite{zhuang2020}, stochastic data-driven representations using  generative adversarial networks \cite{Stengel2020}, reinforcement learning \cite{Novati2021} and with applications ranging from engineering turbulence to geophysics and beyond (see  \cite{brunton2019a} for a recent review). 

One of the great advantage of machine-learning closures is their capability to seamlessly model non-locality in time. Indeed, there is
no a priori reason to expect that the closure terms of a complex system will behave in a Markovian manner, i.e. depend only on the current reduced-order state of the system. In the contrary, Takens embedding theorem
\cite{takens1981detecting}  states that if we observe only a
limited number of the state variables of a system, in principle, we can still
obtain the attractor
of the full system by using delay embedding of the observed state variables. Therefore, it is essential to incorporate memory effects when we model closure terms for turbulent fluid flows.
This approach has found success
in a number of physical applications involving bubble motion and multiphase
flows \cite{wan2018machine, bryngelson2020gaussian}, as well as the
reduced-order modeling of chaotic dynamical systems \cite{Vlachas2018, Wan2018, Vlachas2020}.

On the other hand, machine-learning schemes allows us to
parametrize 
the closure terms using a large number of input variables opening the possibility for  non-local
models in space (see \cite{zhuang2020} for an application to the advaction of a passive scalar). Spatially, non-local models have been advocated for turbulent closures and there is a plethora of related ideas ranging from scale-dependent closures \cite{Porte-Agel2000}, non-local Reynolds Stress models \cite{Hamlington2009}, and fractional-operators closures \cite{Samiee2020}.
Several ideas related to functional neural networks or operator neural networks have shown great promise on this direction \cite{Chen1993, Chen1995}, but not yet utilized in the context of turbulent closure models. 

Beyond local closures deep neural networks have also been used successfully in combination with the underlying governing equations for reconstructing complex fluid flows and identifying flow parameters. Specifically, physics-informed neural networks \cite{Raissi2019} identify the optimal solution (either the flow itself or parameters associated with it) by minimizing an objective function that contains the Navier-Stokes equations, as well as scattered data in space and time. Inclusion of the governing equations significantly improves the behavior of the data-driven scheme, while the representation of the solution in terms of a neural networks circumvents the need for a grid
or spatial discritization scheme.
The method has shown great promise 
for reconstructing fluid flows  given spatio-termporal  measurements \cite{Raissi2020}, as well as recovering macroscopic quantities such as lift or drag for vortex-induced vibration problems \cite{Raissi2019a}.
Previous efforts along this line include the embedding of symmetries such as
Galilean invariance to the neural net predictions for an anisotropic Reynolds
stress tensor \cite{ling2016reynolds, ling2016machine}. 

Our aim in this work is to formulate energy-preserving spatio-temporally non-local turbulent closures which are a priori consistent with the conservation properties of the advection term in Navier-Stokes equations.
Specifically,
we utilize machine learning schemes which represent the effect of the small scales at each spatial location, using as input the large scale features of the
flow in a spatial neighborhood of this location. Past values of the large scale features are also employed as inputs for the turbulent closures in a causal manner. These data-driven schemes are enforced to be consistent with physical constraints expressing the energy
exchanges between resolved and unresolved scales. These constraints follow from the energy-conserving properties of the nonlinear advection operator in Navier-Stokes and have been utilized previously in the context of uncertainty quantification and stochastic closure models \cite{sapsis11a, sapsis_majda_mqg, sapsis_majda_tur, majda2015}. In contrast to previous effort where the full system equation is used as a constraint, assuming perfect knowledge of the equation form and/or parameters (e.g. \cite{Raissi2020}), the formulated constraint in this work expresses a universal property of the advection terms, i.e. that they do no create or destroy kinetic energy of the flow. To improve the stability properties of the computed closures we also employ imitation learning  \cite{bai2018empirical}.

We first formulate the objective function used in the training phase. This step also includes the physical constraint and its derivation using Gauss theorem. We subsequently consider a forced two-dimensional jet flow. We first take into account the invariance of the flow in one direction to derive one-dimensional machine-learned closures using DNS information. As a next step, we apply the method on the computation of two-dimensional turbulent closures that do not rely on this special symmetry. We compare the obtained coarse-scale model with DNS and assess its generalizability properties for different Reynolds numbers, as well as different jet profiles which have not been used in the training phase. We thoroughly examine the role of the physical constraint on the stability properties and accuracy of the coarse-scale equations. In addition, we assess our closure scheme on capturing the evolution of concentration for inertial tracers, such as bubbles and aerosols.  
 
\section{Formulation of energy-preserving closure schemes}
\label{sec:closure}
Our aim is to derive Eulerian, data-driven closure schemes for turbulent fluid flows, as well as for inertial tracers advected by those. 
These closure schemes will not only rely on  DNS training data, but also on the physical constraint that follows from the energy conservation principles that the nonlinear advection terms  satisfy \cite{sapsis11a, majda2015}. The effectiveness of the closure schemes is assessed  by how well the coarse-scale equations can reproduce the mean flow characteristics for problems that reach a statistical equilibrium. Higher order closures may be utilized to improve predictions for higher order statistics such as the flow spectrum. However, in this work we will focus on closures that aim to model the mean flow characteristics.

First, we introduce a spatial-averaging operator that will define the coarse scale version   of the quantities of interest
and their evolution equations. Specifically, we  decompose any field of interest $f$ as
\begin{equation}\label{filtering}
\begin{split}
    f = \overline{f}+f',
\end{split}
\end{equation}
where $\overline{f}$ corresponds to the large-scale component of the quantity
and $f'$ corresponds to the small-scale component. As a result we always have $\overline{f'} = 0 $.

\subsection{Averaged Navier-Stokes equations}

 We consider the Navier-Stokes equations in dimensionless form: 
\begin{equation}\label{NS_equations}
        \frac{D \textbf{u}}{Dt} = -\nabla p + \frac{1}{\text{Re}} \Delta
\textbf{u}  +\nu \nabla^{-4} \textbf{u} +\textbf{F},
\end{equation}

\begin{equation}\label{incompres_cond}
        \nabla \cdot \textbf{u} = 0,
\end{equation}
where $\textbf{u} $ is the velocity field of the fluid, $p$ its
pressure, $\text{Re}$ is the Reynolds number of the flow, $\frac{D}{Dt}$
is the material derivative operator and $\textbf{F}$ denotes an external
forcing term. Parameter $\nu$ is a hypoviscosity
coefficient aiming to remove energy from large scales and maintain the flow
in a turbulent regime. Using the decomposition (\ref{filtering}) into the fluid flow
eqs. (\ref{NS_equations}) and applying the averaging operator we obtain: \begin{equation}\label{NS_equations_filter}
        \partial_t \overline{\textbf{u}} = -\overline{\textbf{u}} \cdot \nabla
\overline{\textbf{u}} -\overline{\textbf{u}' \cdot \nabla \textbf{u}'} -\nabla
\overline{p} + \frac{1}{\text{Re}} \Delta \overline{\textbf{u}}  +\nu \nabla^{-4}
\overline{\textbf{u}} +\overline{\textbf{F}},
\end{equation}

\begin{equation}\label{incompres_cond_filter}
        \nabla \cdot \overline{\textbf{u}} = 0.
\end{equation}
Clearly, the averaged evolution equations do not comprise a closed system
anymore due to the nonlinearity of the advection term. As a result, the
term $\overline{\textbf{u}' \cdot \nabla \textbf{u}'}$ is not defined by
the evolution equations and needs to be parametrized. 
\subsection{Averaged advection equation for inertial tracers}
One can follow a similar process for the advection equation governing the motion of small inertial tracers. In particular for small inertia particles their Lagrangian velocity, $\textbf{v}$,  is a small perturbation of the underlying fluid velocity field \cite{Haller2008}:

\begin{equation}\label{Sapsis_Haller}
        \textbf{v} = \textbf{u} +\epsilon \Bigg( \frac{3 R}{2} -1 \Bigg)
\frac{D \textbf{u}}{Dt} +O(\epsilon^2), \quad \epsilon = \frac{St}{R} \ll
1, \quad R = \frac{2 \rho_f}{\rho_f +2 \rho_p},
\end{equation}
where $\epsilon = \frac{St}{R} \ll 1$ is a parameter representing the importance
of inertial effects, $St$ is the particle or bubble Stokes number,  and $R = \frac{2\varrho_f}{\varrho_f +2\varrho_p}$ is a density ratio with $\varrho_p$ and $\varrho_f$ being the density of the particle or bubble and the flow respectively. 

By introducing $\rho$ as the concentration of tracers
at a particular point, we can write the following
transport equation

\begin{equation}\label{Bubble_Transport}
        \partial_t \rho +\nabla \cdot ( \textbf{v} \rho) = \nu_2 \Delta^4
\rho.
\end{equation}
The right-hand-side of the transport equation represents a hyperviscosity
term. Introducing the decomposition
of eq. (\ref{filtering}) in the evolution eqs. (\ref{Sapsis_Haller}) and
(\ref{Bubble_Transport}), we obtain
\begin{equation}\label{Sapsis_Haller_filter}
        \overline{\textbf{v}} = \overline{\textbf{u}} +\epsilon \Bigg( \frac{3
R}{2} -1 \Bigg) \Big( \partial_t \overline{\textbf{u}} + \overline{\textbf{u}}
\cdot \nabla \overline{\textbf{u}} +\overline{ \textbf{u}' \cdot \nabla \textbf{u}'
}  \Big),
\end{equation}

\begin{equation}\label{Bubble_Transport_filter}
        \partial_t \overline{\rho} +\nabla \cdot ( \overline{\textbf{v} \rho})
+\overline{\nabla \cdot (\textbf{v}' \rho')} = \nu_2 \Delta^4 \overline{\rho}.
\end{equation}
Once again, the closure term $\overline{\nabla \cdot (\textbf{v}' \rho')}$
appears, which requires parametrization. 
\par

\subsection{Data-driven parametrization of the closure terms}
While the full Navier-Stokes equations and the associated advection equations are Markovian and spatially-local, i.e. the evolution of the flow or concentration field in a specific location and time instant depends only on the current time instant and the current neighborhood, this is not necessarily the case for the averaged version of these equations.
In particular, for the averaged equations we typically do
not have access to the full-state information, required to fully describe the evolution of the system. In this case the missing information is the small-scale dynamics.

To overcome this limitation we recall Takens embedding theorem
\cite{takens1981detecting}, which states that if we observe only a
limited number of the state variables of a system, in principle, we can still obtain the attractor
of the full system by using delay embeddings of the observed state variables. In other words, under appropriate conditions there is a map between the delays of the observed state variables and the full state system. Although the theorem itself is several decades old, we can now rely on recently developed data-driven schemes that can implement such mapping  as part of their training process,  enhancing the accuracy of predictions (see e.g. \cite{vlachas2018data, wan2018data}).
To this end, we  parametrize the closure terms with \textit{non-local in time} (but still causal) \textit{} representations, based on Long-Short-Term Memory (LSTM) recurrent neural networks (RNN) and Temporal Convolutional Networks (TCN)  \cite{hochreiter1997long}.
The specific RNN implementation was picked based on its tested ability to
incorporate long-term memory effects of hundreds of time-delays, while simpler
RNN models suffer from vanishing or exploding gradients \cite{bengio1994learning}.

\par
Beyond non-locality in time we also choose to employ non-locality in space. That is, given a
point in space $\mathbf{x}$, we use information from points that
lie in a small neighborhood of $\mathbf{x}$. Clearly, incorporating information from the entirety of the domain is not
only computationally infeasible but also redundant and can lead to stability issues. For this reason we use convolutions
in space to make sure that we incorporate information only from a region
around each point and not from the entire domain. The parameterization is based on a stacked LSTM architecture \cite{graves2013speech}, which utilizes LSTM
layers  with the detail that all input and recurrent transformations are
convolutional. 

As a result, the closure terms are modeled in the following
form:
\begin{equation}\label{Neural_Nets_Models}
\begin{split}
    \overline{\textbf{u}' \cdot \nabla \textbf{u}'} (\mathbf{x},t)&= \mathbb{D}_{\textbf{u}}
\left[{\theta}_1;  \overline{\xi}[\alpha(\mathbf{x}),\chi(t)] \right],\\
    \overline{\nabla \cdot (\textbf{v}' \rho')} (\mathbf{x},t)&= \mathbb{D}_{\rho} \left[\theta_2;
\overline{\zeta}[\alpha(\mathbf{x}),\chi(t)]\right] ,
\end{split}
\end{equation}
where $\overline{\xi}$ and $\overline{\zeta}$ are (averaged) flow features
to be selected, $\alpha(\mathbf{x})$ denotes a pre-selected neighborhood of points around $\mathbf{x}$ over which the averaged state is considered, i.e. $\alpha(\mathbf{x}) = \{ \mathbf{x} , \mathbf{x}+\mathbf{x}_1,\mathbf{x}+\mathbf{x}_2,...,\mathbf{x}+\mathbf{x}_N \} $, and $\chi(t)$ denotes the history of the averaged state backwards from time $t$, i.e. $\chi(t)=\{t,t-\tau_1,...,t-\tau_2,...,t-\tau_N \}$ . The vectors $\theta_1$ and $\theta_2$ denote the hyperparameters of the neural networks and their optimization is
performed empirically. The spatial neighborhood, $\alpha(\mathbf{x})$,  is selected such that if we further increase it, the training error does not significantly
reduce any more.
Note however, that the number of points in space that have to be considered in  $\alpha(\mathbf{x})$
is dependent on the discretization of the domain, i.e., if we increase the resolution of our
model the number of neighborhood points in  $\alpha(\mathbf{x})$ should increase respectively. In the application section we study the effect of spatial discretization. We use a similar approach for the temporal history, $\chi(t)$. 
\subsubsection{Temporal integration}
We point out that our numerical goal is inline prediction.
This means that the neural nets described by eq. (\ref{Neural_Nets_Models})
must be coupled with the evolution eqs. (\ref{NS_equations_filter})-(\ref{Bubble_Transport_filter}).
For a simple forward Euler scheme for temporal integration, this would imply
that by knowing the values of $\overline{\textbf{u}}, \overline{\rho}$ at
time $t$ we can predict the closure terms at time $t$ using eq. (\ref{Neural_Nets_Models})
and use their values to integrate eqs. (\ref{NS_equations_filter})-(\ref{Bubble_Transport_filter})
by one time-step $\delta t$ so that we compute $\overline{u}_1, \overline{\rho}$
at time $t+\delta t$. However, if we want to use a higher-order integration scheme
like a 4th-order explicit Runge-Kutta, we need to evaluate the closure terms
at time $t+\delta t/2$ as well. Since, we do not have access to the required time-history
for such a prediction, we instead integrate in time not by $\delta t$ but
by $2 \delta t$ and thus get a time-integration error of the for $O[ (2 \delta
t)^4 ]$.
\subsection{Physical constraints}
An important feature of our data-driven closure schemes is the requirement to satisfy certain physical principles. Specifically, we utilize the energy flux constraint that the advection term does
not alter the total kinetic energy of the model \cite{sapsis11a, sapsis_majda_mqg}. This constraint, follows from Gauss identity. Specifically, for any scalar fielsd, $\alpha, \beta$, and divergence-free field, $\Phi$, we have from Gauss identity:\begin{equation}
\int_{\Omega} \frac{\partial \alpha}{\partial x_j}\beta \Phi_jd\mathbf x=-\int_\Omega \frac{\partial \beta}{\partial x_j}\alpha \Phi_jd\mathbf x+\int_{\partial \Omega}\alpha \beta \Phi_j n_j d\mathbf x,
\end{equation}  where $n_j$ is the unit vector on the boundary, $\partial \Omega$. Applying the above for $\alpha=\beta=u_k$ and $\Phi_j=u_j$, we obtain the general
three-dimensional constraint:
\begin{equation}
\begin{split}
    \int_{\Omega} \textbf{u} \cdot  (\textbf{u} \cdot \nabla) \textbf{u}
 \mathrm{d}\mathbf{x} =\int_{\partial
\Omega}\mathcal{E}\ \textbf{u} \cdot \textbf{n} \  d\mathbf x, \ \ \ \mathcal{E}=\frac{1}{2}\textbf{u} \cdot \textbf{u}
\end{split}
\label{cons101}
\end{equation}
where $\Omega$ is the domain in which the fluid flow is defined. The above constraint essentially expresses the fact that the nonlinear advection terms do not change the total kinetic energy of the system. In what follows we will consider the case of periodic boundary conditions, where the right hand side in (\ref{cons101}) vanishes. However, the same ideas are applicable for arbitrary boundary conditions. We apply the decomposition (\ref{filtering}) and the spatial averaging operator to this equation 
and obtain:
\begin{equation}
\begin{split}
 \int_{\Omega} \overline{\textbf{u}} \cdot  (\overline{\textbf{u}} \cdot \nabla) \overline{\textbf{u}}
 \mathrm{d}\mathbf{x} + \int_{\Omega} \overline{\textbf{u}} \cdot  \overline{(\textbf{u}' \cdot \nabla) \textbf{u}'}
 \mathrm{d}\mathbf{x}+\int_{\Omega} \overline{{\textbf{u}'} \cdot  (\overline{\textbf{u}}
\cdot \nabla) \textbf{u}'}
 \mathrm{d}\mathbf{x} \\  +\int_{\Omega} \overline{{\textbf{u}'} \cdot  (\textbf{u}'
\cdot \nabla)\overline{ \textbf{u}}}
 \mathrm{d}\mathbf{x}+
 \int_{\Omega}\overline{ {\textbf{u}'} \cdot  (\textbf{u}'
\cdot \nabla) \textbf{u}'}
 \mathrm{d}\mathbf{x}= 0,
\end{split}
\end{equation}
From the last equation we have the \emph{physical constraint} that the closure term $\mathbb{D}_{\textbf{u}}$ must satisfy\begin{align}
\begin{split}
 \int_{\Omega} \overline{\textbf{u}} \cdot  \mathbb{D}_{\textbf{u}}
\left[{\theta}_1;  \overline{\xi}[\alpha(\mathbf{x}),\chi(t)] \right]d\mathbf{x}&=A[\mathbf{u}]
 \triangleq- \int_{\Omega} \overline{\textbf{u}} \cdot  (\overline{\textbf{u}} \cdot
\nabla) \overline{\textbf{u}}
 \mathrm{d}\mathbf{x} -\int_{\Omega} \overline{{\textbf{u}'} \cdot  (\overline{\textbf{u}}
\cdot \nabla) \textbf{u}'}
 \mathrm{d}\mathbf{x} \\ & -\int_{\Omega} \overline{{\textbf{u}'} \cdot  (\textbf{u}'
\cdot \nabla)\overline{ \textbf{u}}}
 \mathrm{d}\mathbf{x}-
 \int_{\Omega}\overline{ {\textbf{u}'} \cdot  (\textbf{u}'
\cdot \nabla) \textbf{u}'}
 \mathrm{d}\mathbf{x,}
 \end{split}
\end{align}  
where $A[\mathbf{u}]$ is a function that depends on the training data and the
discretization. Such a constraint can be added to the training
process in a straightforward manner through the objective function. We emphasize that one could formulate a physical constraint based e.g. on the Navier-Stokes equations directly. However, this assumes exact knowledge of the flow-specifics. This is not the case here since the above constraint expresses a universal property, i.e. that advection terms do not create or destroy energy.
\subsection{Objective function for training}
In terms of the training process itself, we normalize the input and output data
as usually suggested (see e.g. \cite{shalev2014understanding}). The loss function for
this problem is chosen to be the single-step prediction mean square error superimposed with the physical constraint. This can be formulated as
\begin{equation}
\begin{split}
    L(\theta_1) =\int_{\Omega\times T}\Bigg\lvert
\Bigg\lvert   \mathbb{D}_{\textbf{u}}
\left[{\theta}_1;  \overline{\xi} \right]- 
    \overline{(\textbf{u}' \cdot \nabla) \textbf{u}'}
    \Bigg\rvert \Bigg\rvert^2d\mathbf{x}dt\mathbf +\lambda \int_T \Bigg\lvert\int_{\Omega} \overline{\textbf{u}} \cdot  \mathbb{D}_{\textbf{u}}
\left[{\theta}_1;  \overline{\xi} \right]d\mathbf{x}-A[\mathbf{u}]
 \Bigg\rvert dt 
\end{split}
\label{obj_u}
\end{equation}
On the other hand, for the advection equation we have the objective function:
\begin{equation}
\begin{split}
    L(\theta_2) =\int_{\Omega\times T}\Big\lvert
\Big\lvert   \mathbb{D}_{\rho}
\left[\theta_2;
\overline{\zeta}[\alpha(\mathbf{x}),\chi(t)]\right] - \nabla \cdot
\big( \overline{ \textbf{v}' \rho' } \big)  \Big\rvert \Big\rvert^2 d \mathbf{x}dt.
\end{split}
\label{obj_rho}
\end{equation}
An important question is which flow features are important as input for each of the
two models. We examine this issue in detail in the following sections.

\label{sec:5}

\subsection{Imitation learning}
\label{imit_learning}

While we use the single-step prediction for training, our aim is to use these
models for multi-step prediction. Any such predictor introduces errors and these compounding errors change the input distribution for future prediction steps, breaking the train-test i.i.d assumption that is common in supervised learning. Under these circumstances, the error can
be shown to grow exponentially \cite{venkatraman2015improving}. This effect
was observed for our setup as well, with the averaged equations
becoming unstable. To alleviate this problem
we use a version of imitation learning presented in \cite{venkatraman2015improving}, the Data as Demonstrator (DAD) method. Specifically, after we
undergo a round of initial training we make predictions and demand that after
a certain number of time-steps the prediction should again match the training
data. Assuming an error tolerance $\delta_1, \delta_2$ for $\mathbb{D}_{\textbf{u}}$
and $\mathbb{D}_{\rho}$ respectively, we can use Algorithm 1 as shown below.
Note that in this pseudo-algorithm, quantities denoted with $*$ correspond
to predictions of the neural network under the assumptions introduced by the
DAD algorithm. By repeating this process, we create additional training data
which we use together with the original training data to continue
training the model. We follow the exact same approach for the
closure term $ \overline{\nabla \cdot (\textbf{v}' \rho')}$ as well. We repeated
this process 20 times, in order to achieve sufficient robustness for our models (see Algorithm 1).

\begin{algorithm}[H]
\SetAlgoLined
\KwInit{Reference closure terms \{ $\mathbb{D}^{\text{DNS}}_i(t_0) , ...,
\mathbb{D}^{\text{DNS}}_i(t_p)$\} $i = \textbf{u}, \rho$, computed from DNS}

\KwData{ NN architecture, averaging-operator, discretization, $\delta
t$ and input flow features $\overline{\mu}$}
\KwResult{Predicted closure terms \{ $\mathbb{D}^{\text{ML}}_i(t_0) , ...,
\mathbb{D}^{\text{ML}}_i(t_p)$\} $i = \textbf{u}, \rho$.}
 Set $\mu^{\text{Ref}} = \mu^{\text{DNS}}$ and $\mathbb{D}_{\textbf{u}}^{\text{Ref}}
= \mathbb{D}_{\textbf{u}}^{\text{DNS}}$\;
 Train $\mathbb{D}_{\textbf{u}}^{\text{ML}}$ using $\mathbb{D}_{\textbf{u}}^{\text{Ref}}$\;
  
 \For{$i = 1, ..., 20$}{
 \For{$t = 0, \delta t,..., T \delta t$}{
   $s \rightarrow t$\;
  \While{$ \Big\lvert \Big\lvert \mathbb{D}_{\textbf{u}}^{\text{ML}} (s)
-\mathbb{D}_{\textbf{u}}^{\text{DNS}} (s) \Big\rvert \Big\rvert > \delta_1$}{
   Predict $\overline{\textbf{u}}^{\text{ML}} (s+\delta t)$\;
   Predict $\mathbb{D}_{\textbf{u}}^{\text{ML}}(s+\delta t)$\;
   $s \rightarrow s+\delta t$\;
   }
   Compute $ \mathbb{D}_{\textbf{u}}^* (s)$ so that $\overline{\textbf{u}}^{\text{ML}}
(s+\delta t) = \overline{\textbf{u}}^{\text{DNS}} (s+\delta t)$\;
    Set $(\mu^{\text{Ref}},\mathbb{D}_{\textbf{u}}^{\text{Ref}}) = (\mu^{\text{DNS}},\mathbb{D}_{\textbf{u}}^{\text{ML}})
\cup (\mu^* \lvert_s, \mathbb{D}_{\textbf{u}}^{*} (s) )$\;
    }
 Train $\mathbb{D}_{\textbf{u}}^{\text{ML}}$ using $\mathbb{D}_{\textbf{u}}^{\text{Ref}}$\;
 }
 \caption{Training of closure scheme}
\end{algorithm}

\section{Fluid flow setup}
\label{sec:1}
For the validation and assessment of the formulated closures we consider a two-dimensional turbulent jet where bubbles are also advected as passive inertial tracers. Specifically, the velocity field governing the bubbles is different from that of the underlying fluid flow (due to inertia effects), but the bubbles do not affect in any way the underlying fluid flow. 

We setup a turbulent jet that fluctuates around a steady-state jet solution, $\textbf{u}_{jet}$. In its dimensionless form this system of equations can be written as
\begin{equation}\label{NS_equations0}
        \frac{D \textbf{u}}{Dt} = -\nabla p + \frac{1}{\text{Re}} \Delta \textbf{u}  +\nu \nabla^{-4} (\textbf{u}-\textbf{u}_{jet}) +\textbf{F},
\end{equation}

\begin{equation}\label{incompres_cond}
        \nabla \cdot \textbf{u} = 0,
\end{equation}
where $\textbf{u} = (u_1,u_2)$ and $Re=O(10^3)$. The domain is assumed rectangular,  doubly periodic, i.e. $\textbf{x} = (x,y) \in S^{2} =[0,2\pi]\times[0,2\pi]$. For initial conditions, since we desire anisotropy in our flow, we use Gaussian jet structures of the general form
\begin{equation}
        u_{1,jet} = \sum_i A_i \exp \Big[-c_i (y-\beta_i)^2 \Big], \quad u_{2,jet} = 0,
\end{equation}
where $A_i, c_i, \beta_i$ are parameters. The role of the external forcing term, $\textbf{F}$, is twofold: i) it contains a large-scale component to maintain the jet structure, by balancing the dissipation term, and ii) it has a small-scale and small-amplitude component to perturb the flow and trigger instabilities  so that the we enter a turbulent regime. To achieve turbulence we choose a forcing term that acts only on a specific waveband with $6 \leq \lVert \textbf{k} \rVert \leq 7$. Exciting a flow with a forcing localized only in a narrow wavenumber interval is common practice in the turbulence literature \cite{bracco2010reynolds, boffetta2010evidence, mazzino2007scaling, borue1993spectral}. 

Therefore, we adopt a form $\textbf{F} = -\frac{1}{R} \Delta \pmb{u}_{jet} +f$, with $f$ being

\begin{equation}
    f(\textbf{x},t) = \sum_i A_i(t) \cos(\textbf{k} \cdot \textbf{x}+\omega_i),
\end{equation}
where $6 \leq \lvert \textbf{k} \rvert \leq 7$, $A_i(t)$ are random vectors that follow a Gaussian white noise distribution (each one independent from the other) and $\omega_i$ are phases sampled from a uniform distribution over $[0,2 \pi]$. The standard deviation for these amplitudes is set to $0.03$. This ensures that the energy and enstrophy inputs are localized in Fourier space and only a limited range of scales around the forcing is affected by the details of the forcing statistics. Furthermore, such a forcing ensures that the system reaches a jet-like statistical steady state after a transient phase. Due to the small-scale forcing being essentially homogeneous in space we can deduce that the statistical steady state profile is only dependent on $y$ (since our large-scale forcing and initial conditions depend only on $y$). We solve this flow using a spectral method and $256^2$ modes.

\begin{figure}[H]
\centering
        \includegraphics[width=0.90\textwidth]{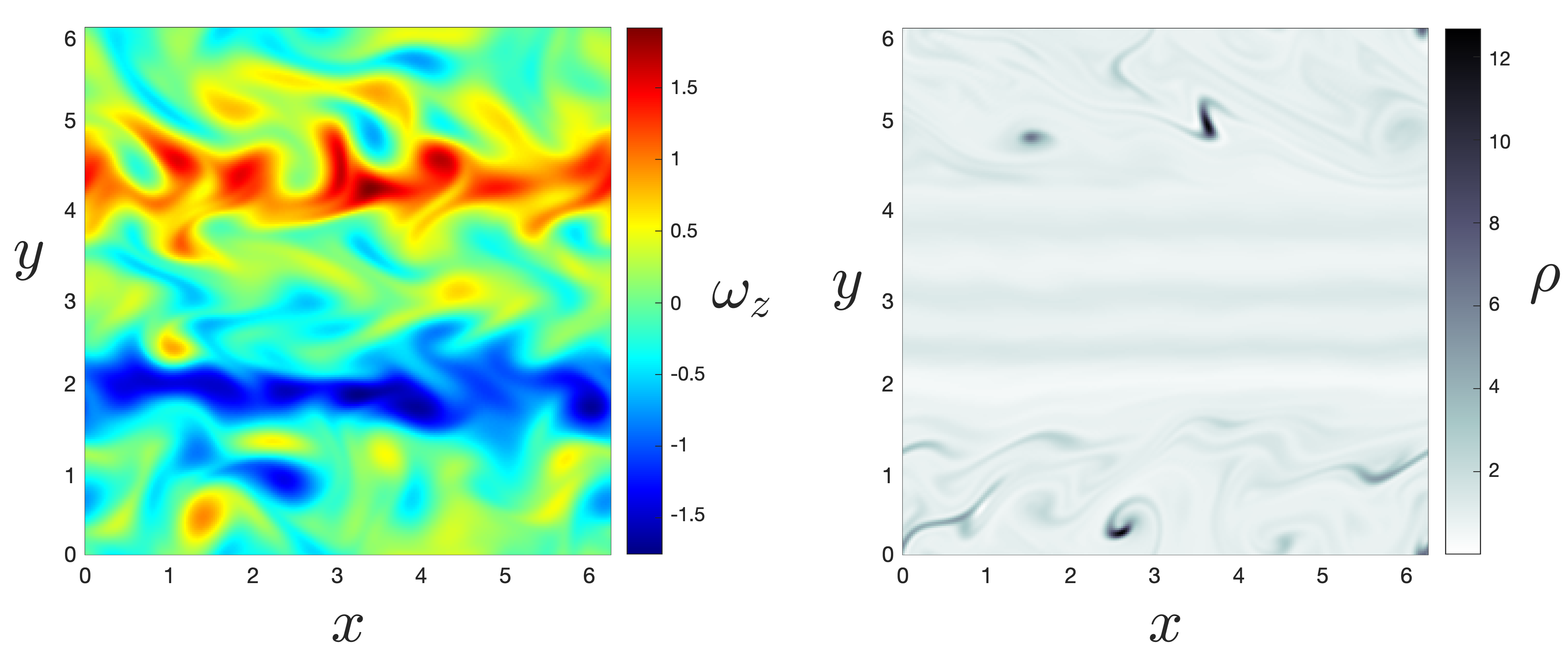}
        \caption{
        Snapshot of vorticity (left) and bubble density field (right) for a bimodal turbulent jet for $Re = 1000$, and bubble parameters $\epsilon = 0.05$ and $R=2$. 
        }
        \label{fig:Flow_Snapshot}
\end{figure}

For the bubbles we use the perturbed advection field (eq. (\ref{Sapsis_Haller})) and the corresponding advection equation (\ref{Bubble_Transport}). For the simulations presented we use the inertial parameters, $\epsilon = 0.05$ and $R = 2$, which correspond to small bubbles.  A typical snapshot of the described flow can be seen in Figure  \ref{fig:Flow_Snapshot}.

\section{Training of the closures}
We study the effectiveness of the proposed closure scheme in two different setups. In the first setup, we take advantage of the translational invariance of the flow in the $x-$direction. This allows us to obtain a closed averaged equation for the $y-$profile of the jet. In the second case we do not rely on this symmetry and obtain closures directly for the two-dimensional flow. We will compare the adopted architectures for both the case of utilizing the energy constraint presented above and not.
\subsection{One-dimensional closures for the jet profile }
\label{feature_selec}
We take advantage of the translational invariance of the flow in the $x-$direction and select the spatial-averaging operator to be integration along the full $x$-direction and local spatial averaging along the $y$-direction:\begin{equation}
\bar f(y)=\frac{1}{2\pi } \iint_{S^2} w_{l}(y'-y)f(x,y') \mathrm{d}x \mathrm{d}y'
\end{equation} where, $w_l(y'-y)=\frac{1}{2l}[H(y'-y-l)-H(y'-y+l)]$ is the piece-wise constant averaging window of length $2l$;   here $l=\frac{2 \pi}{20}$. Applying the averaging operator to the governing equations we obtain the equation for the $y-$profile of the jet.
Note that based on our averaging operator for this case, we have $\bar u_2=0$ and $\bar u_1$ is only a function of $y$.
To this end, we have:\begin{equation}\label{NS_1D_Closure}
\begin{split}
        &\partial_t \overline{u}_1 =   -\mathbb{D}_{{u}_1}  + \frac{1}{\text{Re}}
\partial_y^2 \overline{u}_1  +\nu \nabla^{-4} (\overline{u}_1-u_{1,jet})
+\overline{F}_1,
\end{split}
\end{equation}

\begin{equation}\label{Bubble_1D_Closure}
        \partial_t \overline{\rho} +\partial_y ( \overline{v_2} \hspace{0.1cm} \overline{\rho}) +\mathbb{D}_{\rho}
= \nu_2 \partial_y^4 \overline{\rho}.
\end{equation} 
Therefore, the objective function used for training of the flow closure takes the form:
\begin{equation}\label{1D_Constraint}
\begin{split}
    L(\theta_1) =\int_{\Omega\times T}\Bigg\lvert
\Bigg\lvert   \mathbb{D}_{{u}_1}
\left[{\theta}_1;  \overline{\xi} \right]- 
    \overline{(\textbf{u}' \cdot \nabla) \textbf{u}'}
    \Bigg\rvert \Bigg\rvert^2 \mathrm{d}\mathbf{x} \mathrm{d}t\mathbf +\lambda \int_T \Bigg\lvert\int_{\Omega}
\overline{u}_1 \cdot  \mathbb{D}_{{u}_1}
\left[{\theta}_1;  \overline{\xi} \right] \mathrm{d}\mathbf{x}-A[\mathbf{u}]
 \Bigg\rvert \mathrm{d}t ,
\end{split}
\end{equation}
where $A[\mathbf{u}]=-
 \int_{\Omega}\overline{ {\textbf{u}'} \cdot  (\textbf{u}'
\cdot \nabla) \textbf{u}'}
 \mathrm{d}\mathbf{x}$.
On the other hand, the objective function for the density field closure takes the form:
\begin{equation}
L(\theta_2) =\int_{\Omega\times T}\Big\lvert
\Big\lvert   \mathbb{D}_{\rho}
\left[\theta_2;
\overline{\zeta}[\alpha(\mathbf{x}),\chi(t)]\right] - \nabla \cdot
\big( \overline{ \textbf{v}' \rho' } \big)  \Big\rvert \Big\rvert^2 d \mathbf{x}dt.
\end{equation}
The  neighborhood $\alpha(y)$ is selected to have five nodes in total: \begin{displaymath}
\alpha(y)=\{y+m\delta y\}, \ \delta y=2 \pi /80, \ \  m=-2,-1,...,2,
\end{displaymath} while the temporal horizon in the past is selected as\begin{displaymath}
\chi(t)=\{t-m\delta \tau\}, \delta \tau=10^{-2}, \ m=1,2,...,12.
\end{displaymath}
Both the spatial extent of the neighborhood and the memory are chosen as the threshold values above which any further increase does not result significant difference in the training and validation errors.

\subsubsection{Neural network architecture}
\begin{figure}[H]
    \centering
    \subfloat[]{{\includegraphics[width=0.40\textwidth]{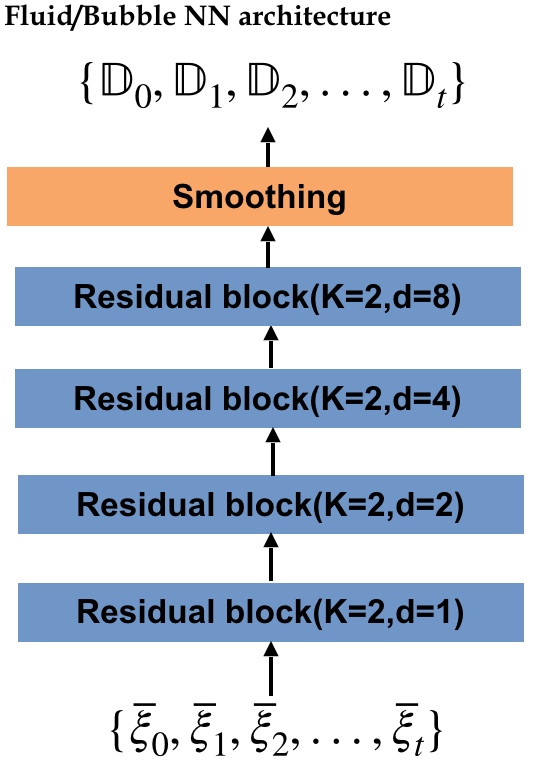} }}%
    \qquad
    \subfloat[]{{\includegraphics[width=0.40\textwidth]{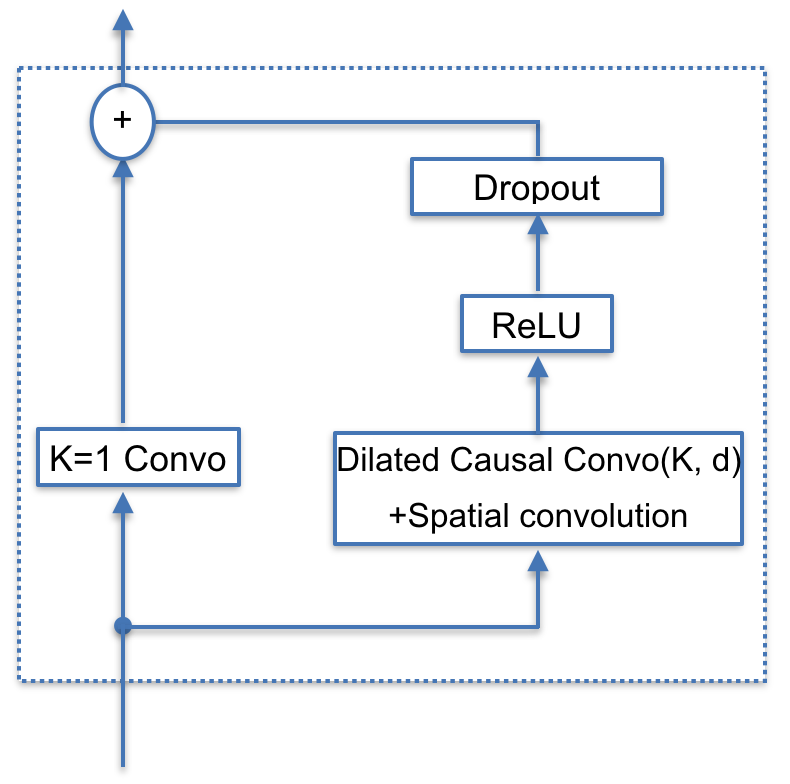} }}%
    \qquad
    \subfloat[]{{\includegraphics[width=0.45\textwidth]{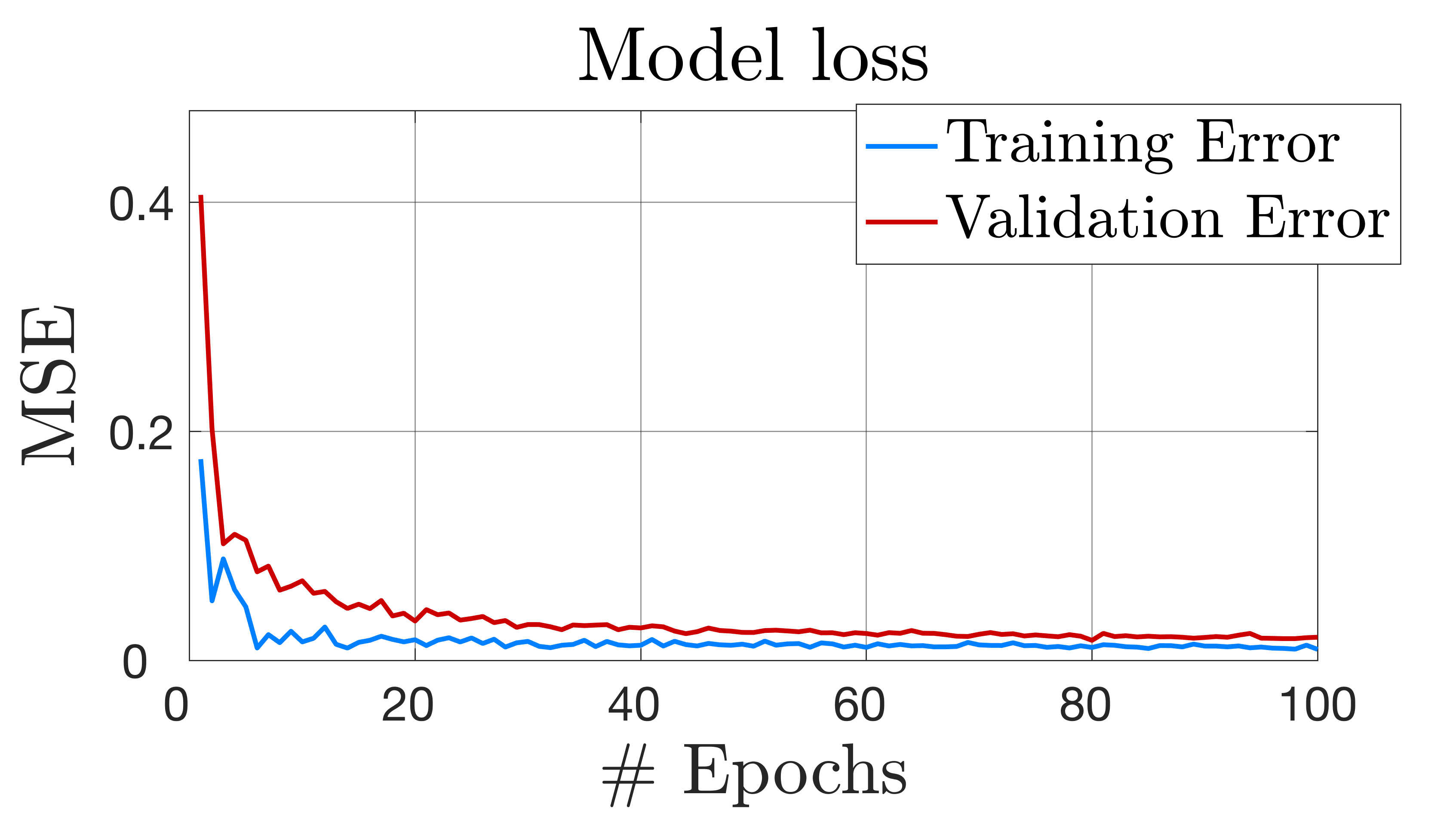} }}%
    \subfloat[]{{\includegraphics[width=0.45\textwidth]{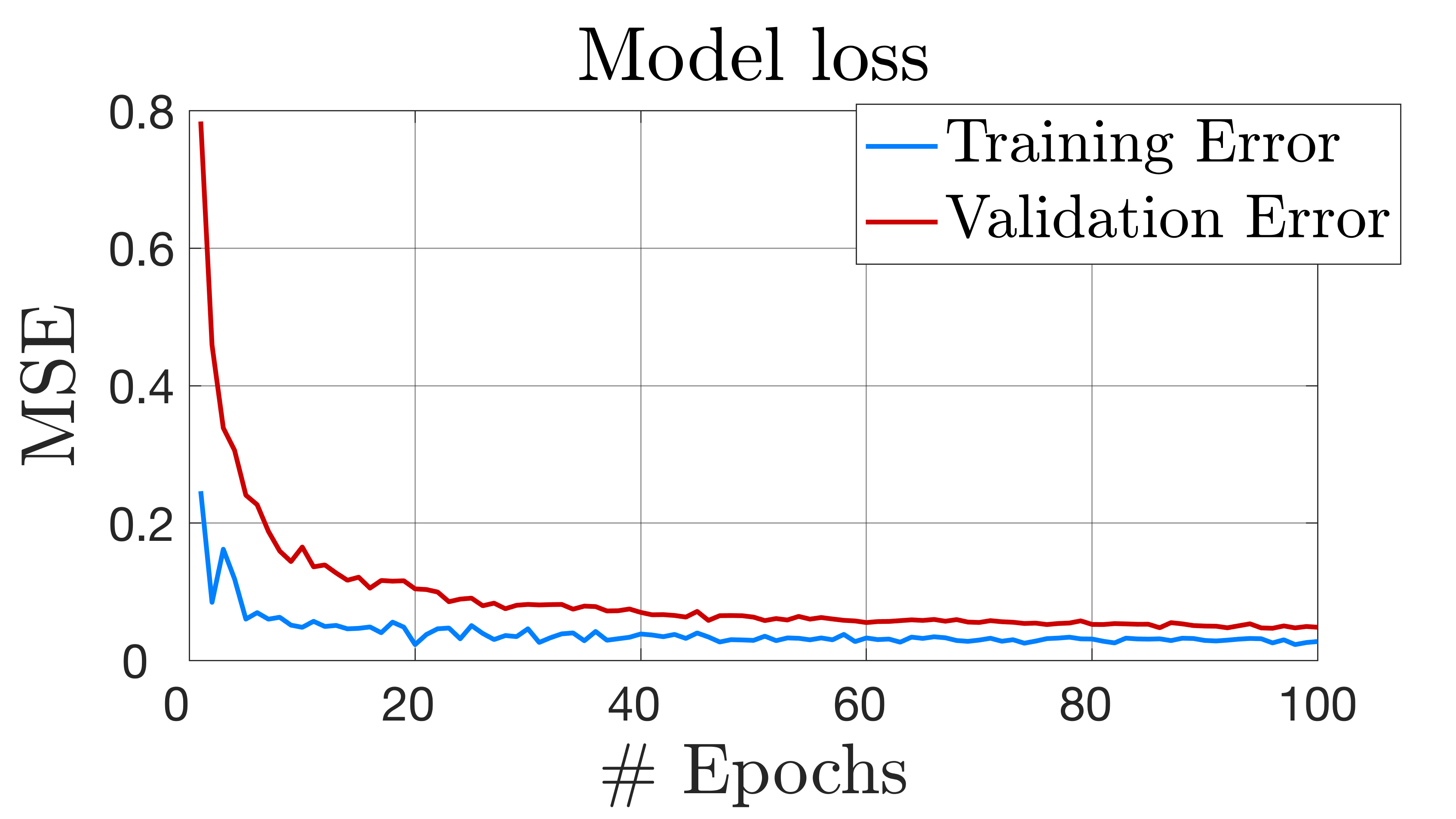} }}%
    \caption{(a) Architecture of the TCN neural network for parametrizing the term $\partial_y(\overline{u_1' u_2'})$ and  $\partial_y ( \overline{ v_2' \rho'})$. (b) Inner architecture of a residual block. (d) Mean square training-error (solid line) and validation error (dashed line) for $\mathbb{D}_{u}$. (e) Mean square training-error (solid line) and validation error (dashed line) for $\mathbb{D}_{\rho}$.}%
    \label{fig:TCN_Architecture}%
\end{figure}

We assess two different architectures for our closure scheme. In the first case, we represent the flow closure with 3 convolutional-LSTM layers and the density closure with 4 convolutional-LSTM layers (16 time-delays). Further increase of the number of layers does not offer any significant improvement in the training and validation errors. The adopted architectures are presented in Figure \ref{fig:example} (a-b). For the second machine learning architecture we use 4-layer  temporal convolutions to model the memory terms of our closure for both the flow and the density fields.  The architecture in this case is depicted in Figure \ref{fig:TCN_Architecture}(a-b). An important difference between the two architectures that is worth emphasizing is associated with their computational cost. Specifically, in the LSTM architecture we have a memory term
that is updated at each time-step and to this end, LSTM needs to only operate on the flow
features at each time-step. On the other hand, TCN layers operate on the entire included time-history
making them more computationally expensive.

\subsubsection{Feature selection}
The selection of the flow features that are used as inputs for the data-driven closures is done numerically by testing different combinations of basic flow features. We eventually choose  the combination that minimizes the training and the validation error. It is important to emphasize that if we rely only on the training error we run the risk of over-fitting. We carry out this process individually for each of the closure terms for the Navier-Stokes equation and the transport equation. 

For the closure term $\mathbb{D}_{\textbf{u}}$ we select as possible flow features the quantities: $\overline{u}_1, \partial_t {\overline{u}}_1, \partial_y (\overline{u_1' u_2'})$. Table 1 summarizes performance for different combinations over 100 periods for the TCN and LSTM architectures with the physical constraint.
We observe that the single most important feature is the ingredients of the material derivative,  $D \overline{u}_1/Dt$. Also, we note that the best validation result is achieved by employing all the considered flow features. 
\begin{table}[H]
\centering
\caption{Feature selection for the one-dimensional closure of the Navier-Stokes.}
\begin{tabular}{ |P{3cm}||P{1cm}|P{2cm}|P{2cm}|P{2cm}|P{2cm}|  }
 \hline
 \multicolumn{6}{|c|}{$\xi$ feature selection} \\
 \hline
 \multicolumn{2}{|c|}{  }&\multicolumn{2}{|c|}{cTCN  }&\multicolumn{2}{|c|}{cLSTM  } \\
 \hline
 Features & Dim & Train-MSE & Val-MSE  & Train-MSE & Val-MSE \\
 \hline
 $\overline{u}_1$   & 1    & 0.094 &   1.712 & 0.102 &   1.501  \\
 $\partial_t \overline{u}_1$&   1  & 0.037   & 0.535 & 0.041 &   0.501 \\
 $\partial_y ( \overline{u_1'u_2'} )$ & 1  & 0.028 &  0.144 & 0.033 &   0.139\\
 $ \overline{u}_1, \partial_t \overline{u}_1 $    & 2 & 0.042 & 0.418 & 0.056 &   0.511\\
 $ \overline{u}_1, \partial_y ( \overline{u_1'u_2'} ) $ &   2  & 0.023 &
0.159 & 0.028 &   0.157\\
 $ \partial_t \overline{u}_1, \partial_y (\overline{u_1'u_2'}) $ & 2  & 0.021
  & 0.092 & 0.026 &   0.085\\
 $ \overline{u}_1, \partial_t \overline{u}_1, \partial_y ( \overline{u_1'u_2'}
) $ & 3  & 0.021 & 0.029 & 0.025 &   0.032\\
 \hline
\end{tabular}
\end{table}

\par
\begin{figure}[H]
    \centering
\subfloat[]{{\includegraphics[width=0.40\textwidth]{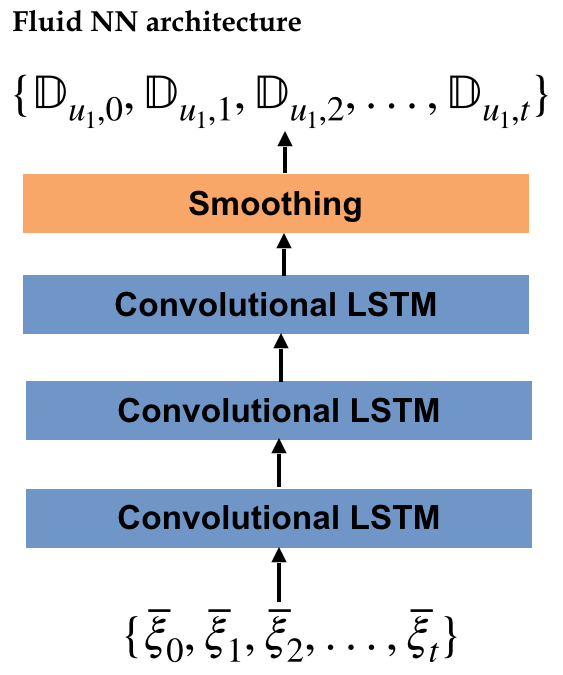} }}%
  \qquad
    \subfloat[]{{\includegraphics[width=0.40\textwidth]{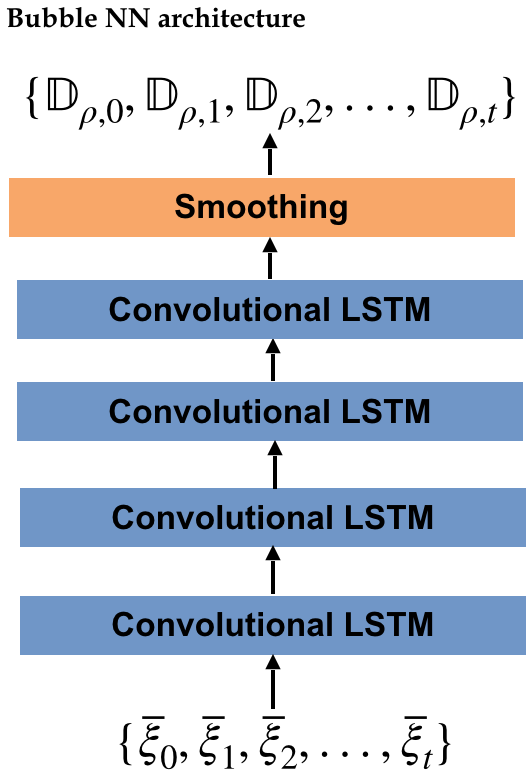} }}%
   
    \subfloat[]{{\includegraphics[width=0.45\textwidth]{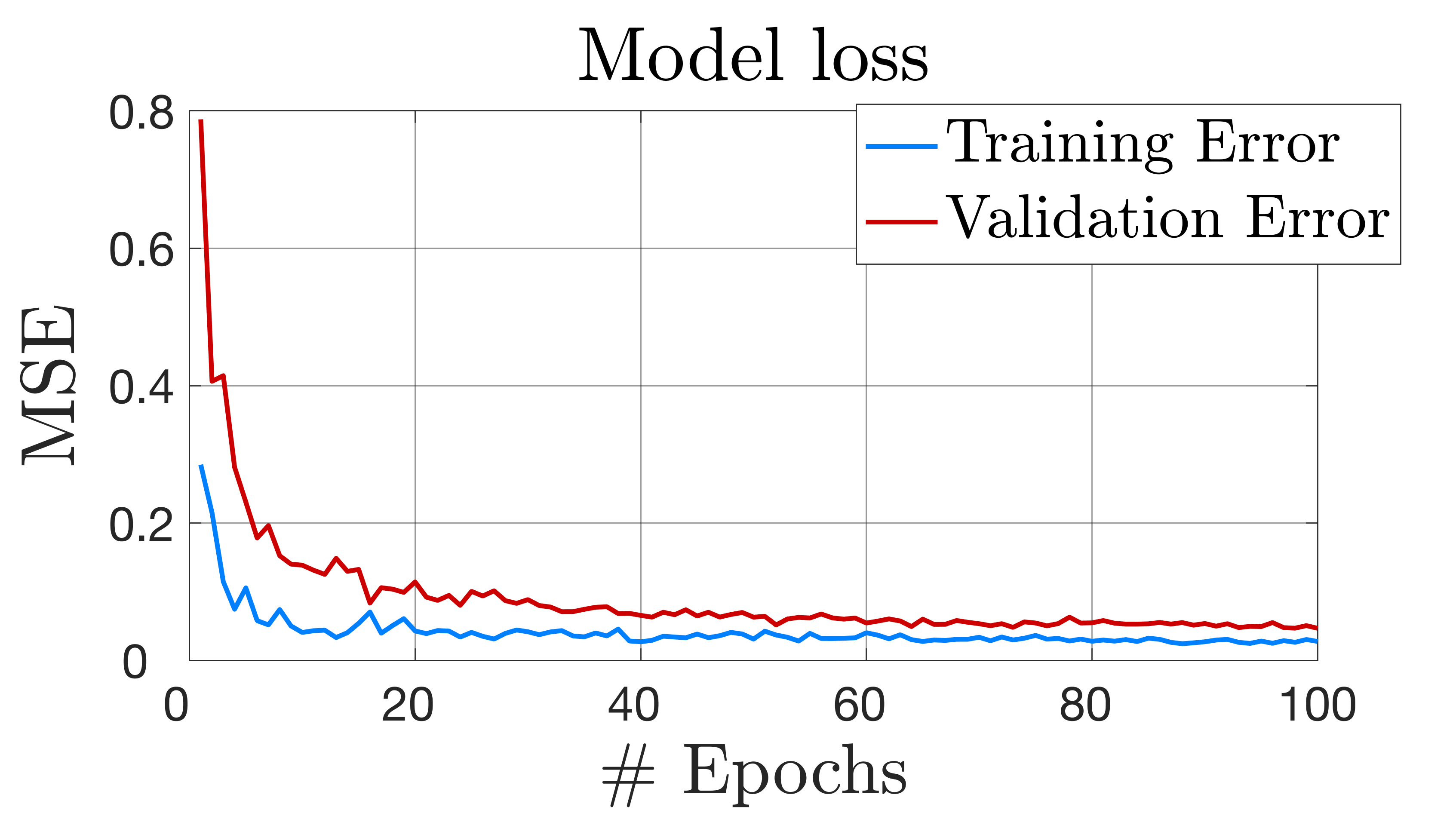} }}%
    \subfloat[]{{\includegraphics[width=0.45\textwidth]{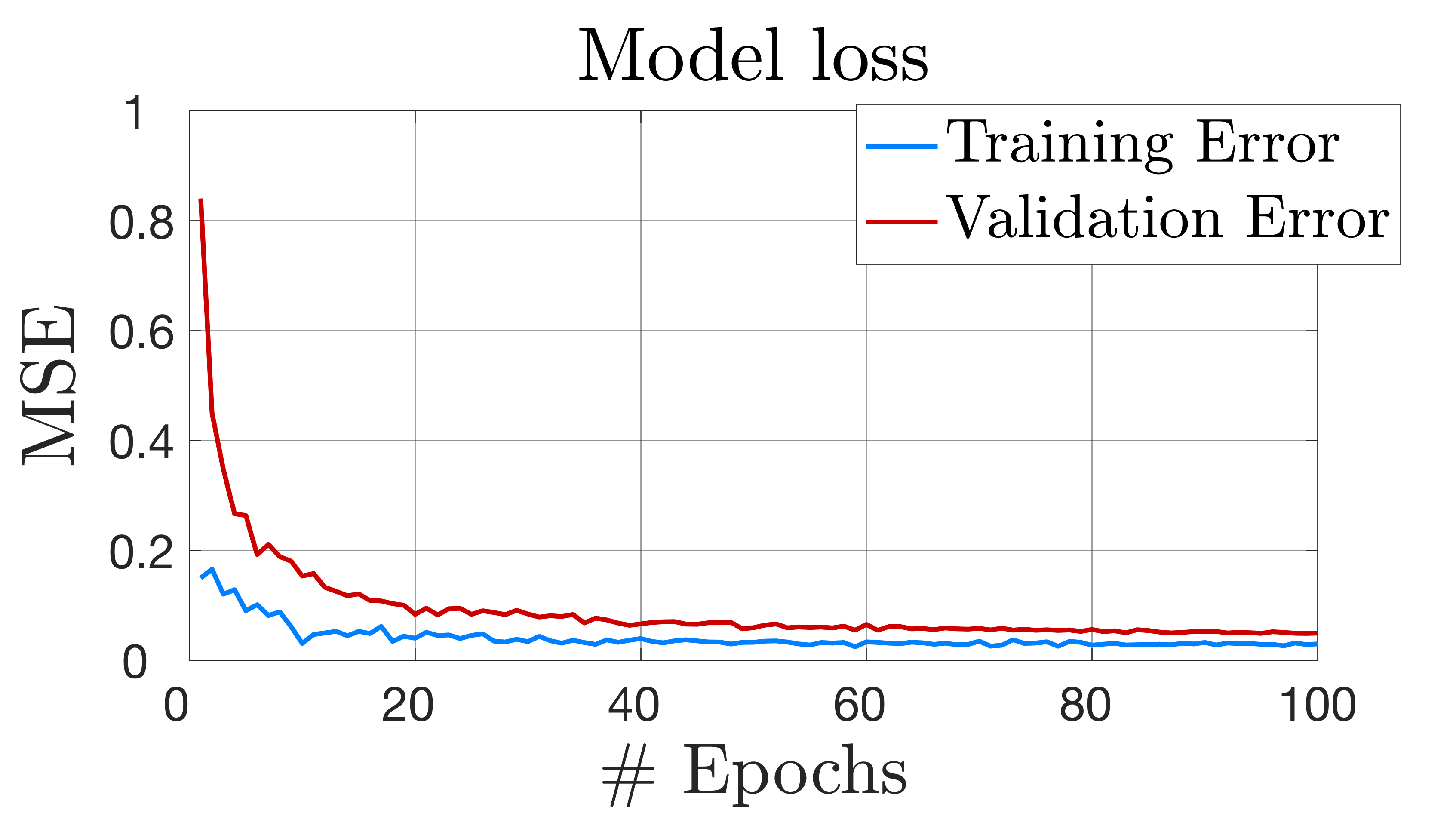} }}%
    \caption{(a) Architecture of the LSTM neural network for parametrizing the term $\mathbb{D}_t = \partial_y(\overline{u_1' u_2'})$. (b) Architecture of the LSTM neural network for parametrizing the term $\mathbb{D}_t = \partial_y ( \overline{ v_2' \rho'})$.  (c) Mean square training-error (solid line) and validation error (dashed line) for $\mathbb{D}_{u}$. (d) Mean square training-error (solid line) and validation error (dashed line) for $\mathbb{D}_{\rho}$.}%
    \label{fig:example}%
\end{figure}

For the closure of the transport equation we carry out the same process in Table 2, where we present the training and validation error over 100 periods. We observe that the single most important feature is $\overline{\rho}$. Based on the mean-square error (both training and validation) we choose the combination of $\overline{u}, \overline{\rho}, \partial_t \overline{u}, \partial_y (\overline{\rho'v'})$. In Figure \ref{fig:example} we present the value of both validation and training error with respect to the number of epochs. These are quite similar, hinting towards generalizability of the predictions.

\begin{table}[H]
\centering
\caption{Feature selection for closure of bubble transport equation.}
\begin{tabular}{ |P{3cm}||P{1cm}|P{2cm}|P{2cm}|P{2cm}|P{2cm}|  }
 \hline
 \multicolumn{6}{|c|}{$\zeta$ feature selection} \\
  \hline
 \multicolumn{2}{|c|}{  }&\multicolumn{2}{|c|}{cTCN }&\multicolumn{2}{|c|}{cLSTM  } \\
 \hline
 Features & Dim & Train-MSE & Val-MSE & Train-MSE & Val-MSE \\
 \hline
 $\overline{\rho}$   & 1    & 0.109 &   0.150 & 0.123 &   0.171\\
 $\overline{\pmb{v}}$ &   2  & 0.603   & 0.673 & 0.592 &   0.625\\
 $\overline{\pmb{v}}, \overline{\rho}$ & 3  & 0.081 &  0.090 & 0.094 &   0.101\\
 $\overline{\pmb{v}}, \overline{\rho}, \partial_t \overline{\pmb{v}}, \partial_t \overline{\rho} $    & 6 & 0.058 & 0.060 & 0.061 &   0.064\\
 $\overline{\pmb{v}}, \overline{\rho}, \partial_t \overline{\pmb{v}}, \partial_y (\overline{\rho'v_2'})  $ &   6  & 0.028 & 0.039 & 0.042 &   0.088\\
 $\overline{\pmb{v}}, \overline{\rho}, \partial_t \overline{\rho}, \partial_y ( \overline{\rho'v_2'} ) $ & 5  & 0.027   & 0.036 & 0.035 &   0.049\\
 $\overline{\pmb{v}}, \overline{\rho}. \partial_t \overline{\pmb{v}}, \partial_t \overline{\rho}, \partial_y ( \overline{\rho'v_2'} ) $ & 7  & 0.025 & 0.031 & 0.033 &   0.044\\
 \hline
\end{tabular}
\end{table}

\subsection{Two-dimensional closures}
\label{feature_selec2}

In this case our averaging operator was chosen to be local in both dimensions, $x$ and $y$:\begin{equation}
\bar f(x,y)=\frac{1}{2\pi } \iint_{S^2} w_{l_x}(y'-y)w_{l_y}(y'-y)f(x,y')dxdy',
\label{smooth2d}
\end{equation} where $l_x = l_y=\frac{2 \pi}{12}$ are the averaging windows in the $x$ and $y$ directions respectively. As objective functions we used equations (\ref{obj_u}) and (\ref{obj_rho}). The spatial and temporal neighborhood used in the closures are chosen as: \begin{displaymath}
\alpha(\mathbf y)=\{\mathbf y+m_1\delta \mathbf x+m_2\delta \mathbf y\}, \ \delta x=2 \pi /48, \delta y=2 \pi /48  \ \text{\ and} \ \  m_{1},m_2=-2,-1,...,2,
\end{displaymath}
\begin{displaymath}
\chi(t)=\{t-m\delta \tau\}, \delta \tau= 1/100, \ m=1,2,...,12.
\end{displaymath}
Similarly with the one-dimensional closures, these numbers are based on the fact that further increase
did not result significant difference in the training and validation errors.
We employ the same neural network architectures that we used in the previous section.
\subsubsection{Feature selection}
For the closure term $\mathbb{D}_{\textbf{u}}$ (corresponding to the fluid flow) we try as possible flow features the quantities $\overline{\pmb{u}}, \partial_t {\overline{\pmb{u}}}, \mathbb{D}_{\textbf{u}}$. Results are shown for the case where the constraint is adopted (cTCN and cLSTM) in Table 3 in terms of training and validation errors.
In this case, we observe that the single most important feature is the history of the Reynolds stresses. Furthermore, we see that the optimal combination consists of all the examined features.  
\par
For the closure of the transport equation we carry out the same process in Table 4. We observe that the single most important features seems to be $\overline{\rho}$, similarly with the one-dimensional case. For the results that follow we choose the combination $\overline{\textbf{v}}, \overline{\rho}, \partial_t \overline{\textbf{v}}, \partial_t \overline{\rho}, \mathbb{D}_{\rho}$, which results in the minimum validation and testing errors. 


\begin{table}[H]
\centering
\caption{Feature selection for closure of Navier-Stokes.}
\begin{tabular}{ |P{3cm}||P{1cm}|P{2cm}|P{2cm}|P{2cm}|P{2cm}|  }
 \hline
 \multicolumn{6}{|c|}{$\xi$ feature selection} \\
 \hline
 \multicolumn{2}{|c|}{} & \multicolumn{2}{|c|}{cTCN} & \multicolumn{2}{|c|}{cLSTM} \\
 \hline
 Features & Dim & Train-MSE & Val-MSE & Train-MSE & Val-MSE \\
 \hline
 $\overline{\pmb{u}} = ( \overline{u_1}, \overline{u_2} )$   & 2    & 0.235 &   0.521 & 0.258 &   0.572\\
 $\partial_t \overline{ \pmb{u}} = ( \partial_t \overline{u_1}, \partial_t \overline{u_2} )$&   2  & 0.098   & 0.480 & 0.112 &   0.388\\
 $\mathbb{D}_{\textbf{u}}$ & 2  & 0.081 &  0.094 & 0.092 &   0.114\\
 $ \overline{\pmb{u}}, \partial_t \overline{\pmb{u}} $    & 4 & 0.069 & 0.485 & 0.100 &   0.522\\
 $ \overline{\pmb{u}}, \mathbb{D}_{\textbf{u}} $ &   4  & 0.048 & 0.082 & 0.067 &   0.094\\
 $ \partial_t \overline{\pmb{u}}, \mathbb{D}_{\textbf{u}} $ & 4  & 0.027   & 0.048 & 0.034 &   0.063\\
 $ \overline{\pmb{u}}, \partial_t \overline{\pmb{u}}, \mathbb{D}_{\textbf{u}} $ & 6  & 0.020 & 0.039 & 0.027 &   0.055\\
 \hline
\end{tabular}
\end{table}

\begin{table}[H]
\centering
\caption{Feature selection for closure of bubble transport equation.}
\begin{tabular}{ |P{3cm}||P{1cm}|P{2cm}|P{2cm}|P{2cm}|P{2cm}|  }
 \hline
 \multicolumn{6}{|c|}{$\zeta$ feature selection} \\
 \hline
 \multicolumn{2}{|c|}{} & \multicolumn{2}{|c|}{cTCN} & \multicolumn{2}{|c|}{cLSTM} \\
 \hline
 Features & Dim & Train-MSE & Val-MSE & Train-MSE & Val-MSE \\
 \hline
 $\overline{\rho}$   & 1    & 0.199 &   0.398 & 0.228 &   0.451\\
 $\overline{\pmb{v}} = (\overline{v_1}, \overline{v_2})$ &   2  & 0.320   & 0.591 & 0.318 &   0.515\\
 $\overline{\pmb{v}}, \overline{\rho}$ & 3  & 0.141 &  0.386 & 0.162 &   0.404\\
 $\overline{\pmb{v}}, \overline{\rho}, \partial_t \overline{\pmb{v}} $    & 5 & 0.085 & 0.176 & 0.087 &   0.192\\
 $\overline{\pmb{v}}, \overline{\rho}, \partial_t \overline{\rho}, \mathbb{D}_{\rho}  $ &   5  & 0.051 & 0.091 & 0.061 &   0.125\\
 $\overline{\pmb{v}}, \partial_t \overline{\rho}, \partial_t  \overline{\pmb{v}}, \mathbb{D}_{\rho}  $ & 6  & 0.030   & 0.049 & 0.038 &   0.068\\
 $\overline{\pmb{v}}, \overline{\rho}, \partial_t \overline{\rho}, \partial_t  \overline{\pmb{v}}, \mathbb{D}_{\rho}  $ & 7  & 0.016 & 0.032 & 0.031 &   0.047\\
 \hline
\end{tabular}
\end{table}


\section{Validation and generalizability for one-dimensional closures}
\label{sec:7}

To showcase the generalizability properties of the obtained closures we train on unimodal jets and we test on bimodal ones. We mention again that the averaged
model is one-dimensional and we use $80$ points in space to simulate
it. We compare the results of the averaged model with the predictions of
the two-dimensional reference solutions that we computed using a spectral method and $256^2$
modes. Each training case contains data in the time-interval $T=
[200, 600]$. 

\subsection{Validation on unimodal jets}\label{uni_modal1}
\label{sec:8}
\begin{figure}[bt]
\centering
        \includegraphics[width=0.75\textwidth]{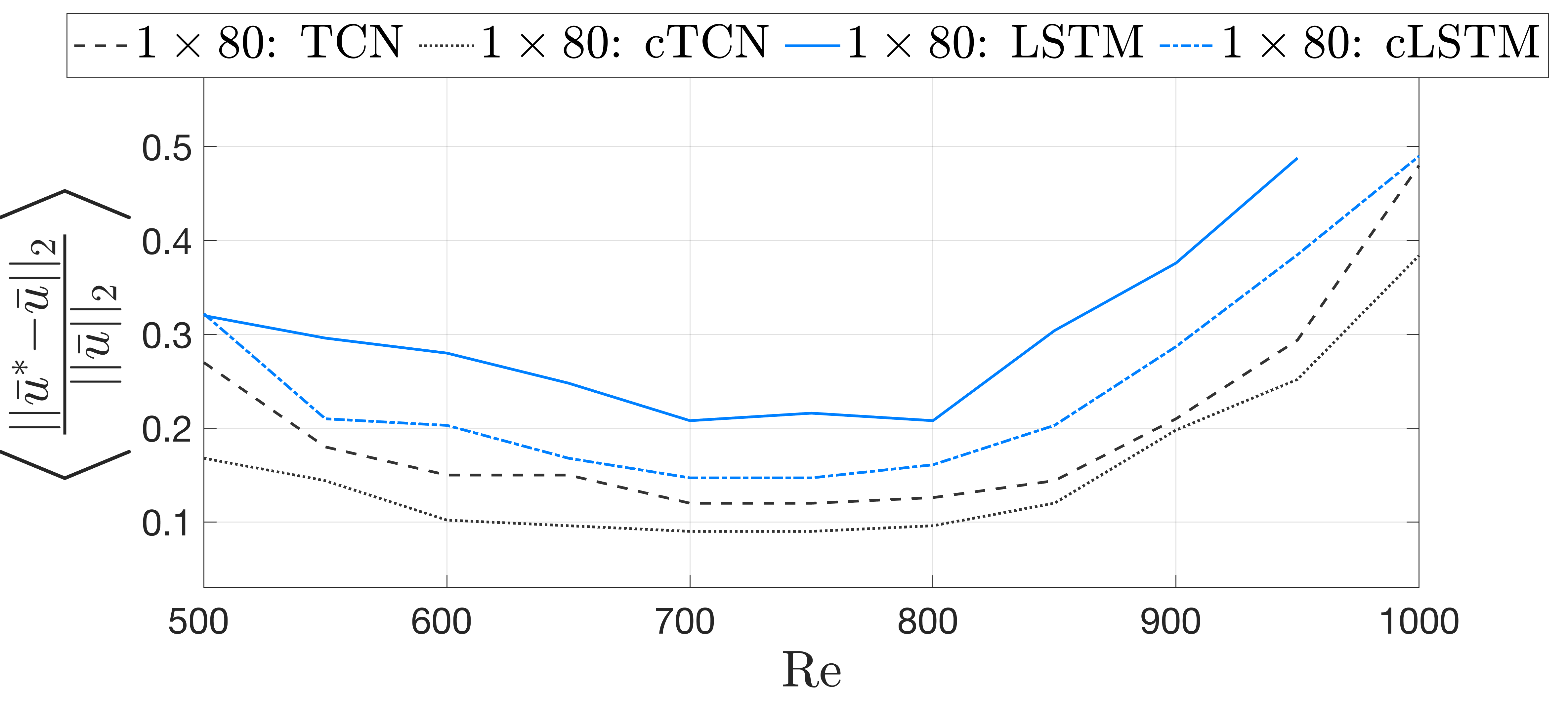}
        \caption{
        Normalized error (\ref{error1d}) for  one-dimensional closure models using TCN, LSTM and and their constrained versions, cTCN and cLSTM for unimodal jets. Training data includes unimodal jets with $Re \in \{650,750,850\}$.
        }
        \label{fig:L2_Error_unimodal1D}
\end{figure}

The unperturbed  jet profile is chosen as,

\begin{equation}
        u_{1,jet} = \exp \Big[-2 (y-\pi)^2 \Big], \quad u_{2,jet} = 0.
\end{equation}
We train four different models on unimodal jets of $Re \in \{650,750,850\}$. We use LSTM and TCN architectures with and without enforcing the physical constraint of eq. (\ref{1D_Constraint}). In Figure \ref{fig:L2_Error_unimodal1D}, we present the time- and $y-$ averaged mean-square error between the $x-$averaged profile of the DNS, $\bar u^*$ and the coarse scale model, $\bar u$:
\begin{equation}
    ||\bar u^*-\bar u||^2_2=\frac{1}{2\pi T}\int_0^{2\pi}\int_{t_0}^{t_0+T}(\bar u^*(y,t)-\bar u(y,t))^2dydt.
    \label{error1d}
\end{equation}
We observe that the TCN models clearly outperform the LSTM based closures. Moreover, training using the objective function that includes the physical constraint (eq. (\ref{1D_Constraint})) for the advective terms (cTCN and cLSTM) improves significantly the testing results for the two architectures by $23\%$ and $25\%$, respectively (Table 5). This improvement comes at no additional cost in terms of data, but only using the physical constraint associated with the advection terms, which does not depend on the knowledge of any physical quantity of the flow or any other system-specific information. 

\begin{figure}[bt]
\centering

    \subfloat[]{{\includegraphics[width=0.45\textwidth]{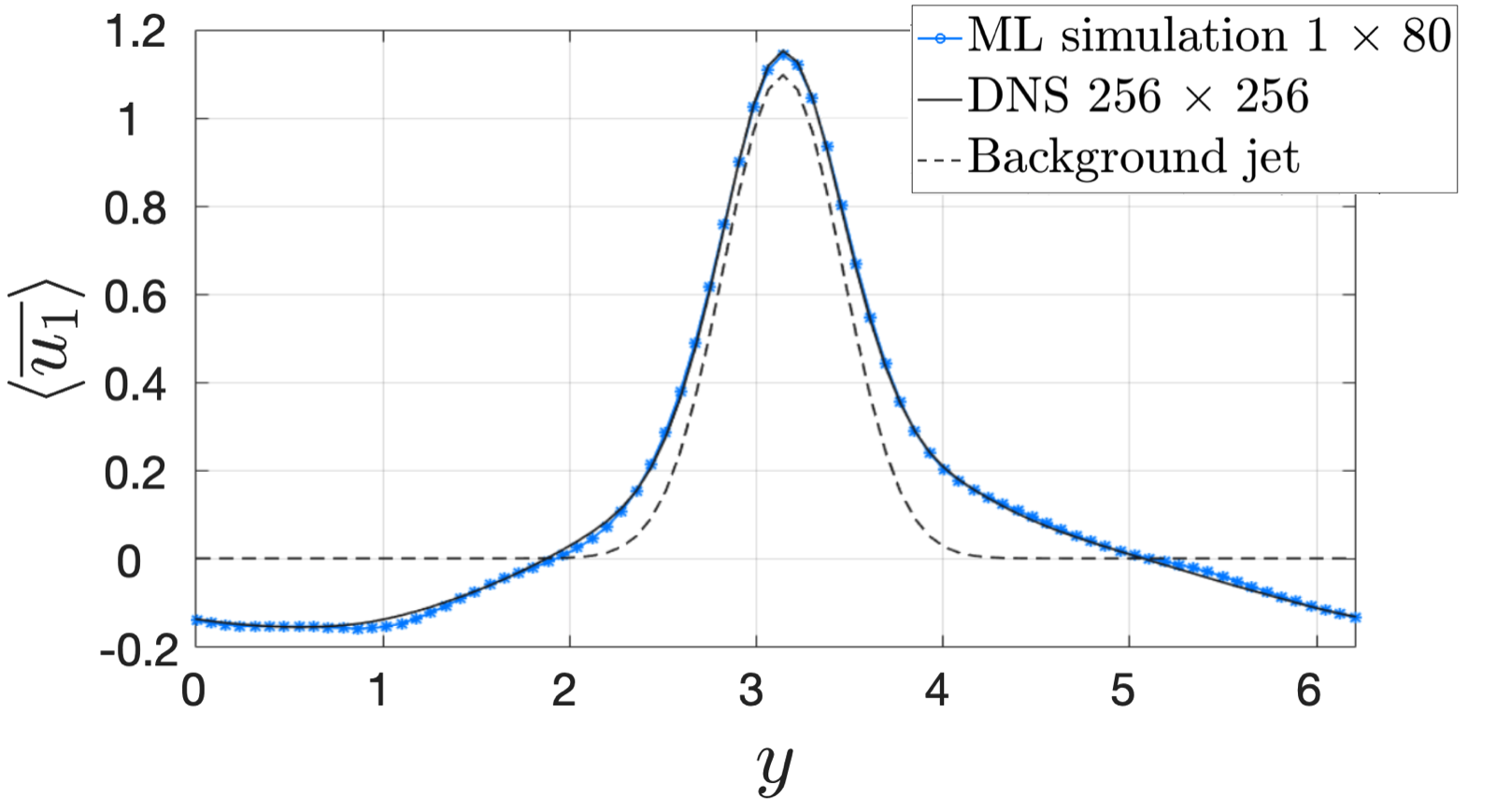}}}
    \subfloat[]{{\includegraphics[width=0.45\textwidth]{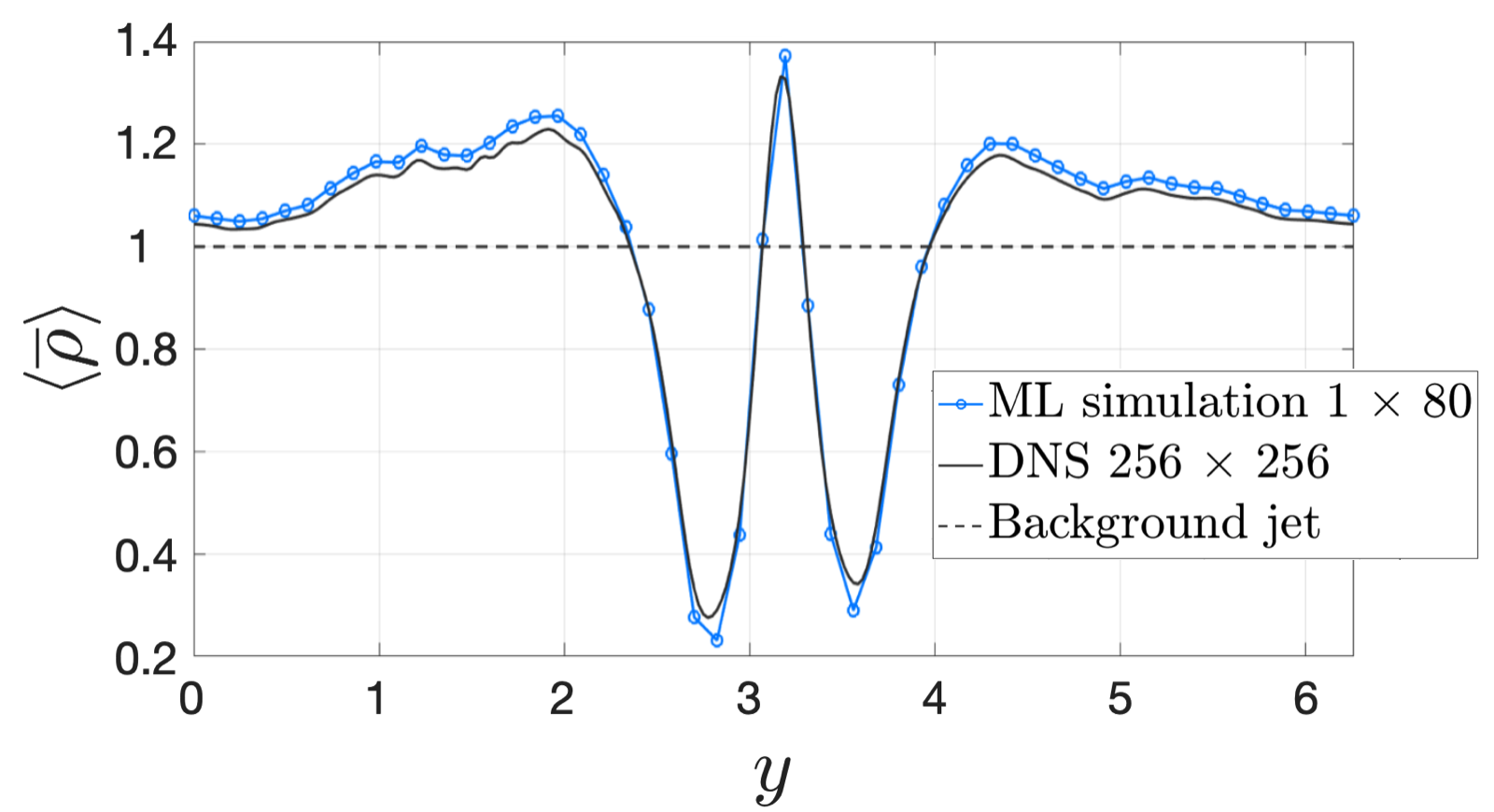}}}

        \caption{
        Time-averaged profile of $\overline{u}_1$ (left) and $\bar \rho$ (right) for the one-dimensional cTCN closure model (blue line) and the DNS (black line). The shape of the jet that is imposed by the large-scale forcing is depicted with dashed line. The simulation corresponds to $Re = 1000$ while training included $Re \in
\{650,750,850\}$. 
        }
        \label{Unimodal_flow1D}
\end{figure}

In Figure \ref{Unimodal_flow1D} we present additional results for the cTCN model (best performer). We showcase results both for the time-averaged profile for the fluid velocity and for the bubble distribution for $Re = 1000$. Comparisons are made between the time-averaged results that the one-dimensional closure scheme produces and the time- and $x-$ averaged results of the two-dimensional reference solution. Specifically, the time-averaged jet-profile is computed as
\begin{equation}
        \langle \overline{u}_1 \rangle (y)=
        \frac{1}{T}\int_{t_0}^{t_0+T} \overline{u}_1^{} (y,t)dt
        ,
\end{equation}
where $\bar u_1^{}$ is the $x-$averaged  reference solution and the temporal averaging parameters are chosen as $t_0=200$ and $T=400$; note that we omit the first transient part of the simulation.

The results appear to be in very good agreement showcasing that our closure scheme is able to predict the statistical steady state of the flow. We can also observe the Rayleigh instability that the initial jet profile (dashed line) undergoes due to the excitation by the external forcing. We note that the slight asymmetry that the velocity profile exhibits is due to a minor inhomogeneity of the forcing term along the $y$ direction. We apply the same operation described above to compute $\langle \overline{\rho} \rangle$ from both the reference solution and the data-based closure scheme (Figure \ref{Unimodal_flow1D}(b)). Again, we obtain very good agreement between the machine learning approach and the reference solution, which has a non-trivial form, as the bubbles seem to cluster around the core of the jet and be repelled from the adjacent areas of the jet core.

\subsection{Testing generalizability on bimodal jets}
\label{sec:9}

\begin{figure}[tb]
\centering
        \includegraphics[width=0.75\textwidth]{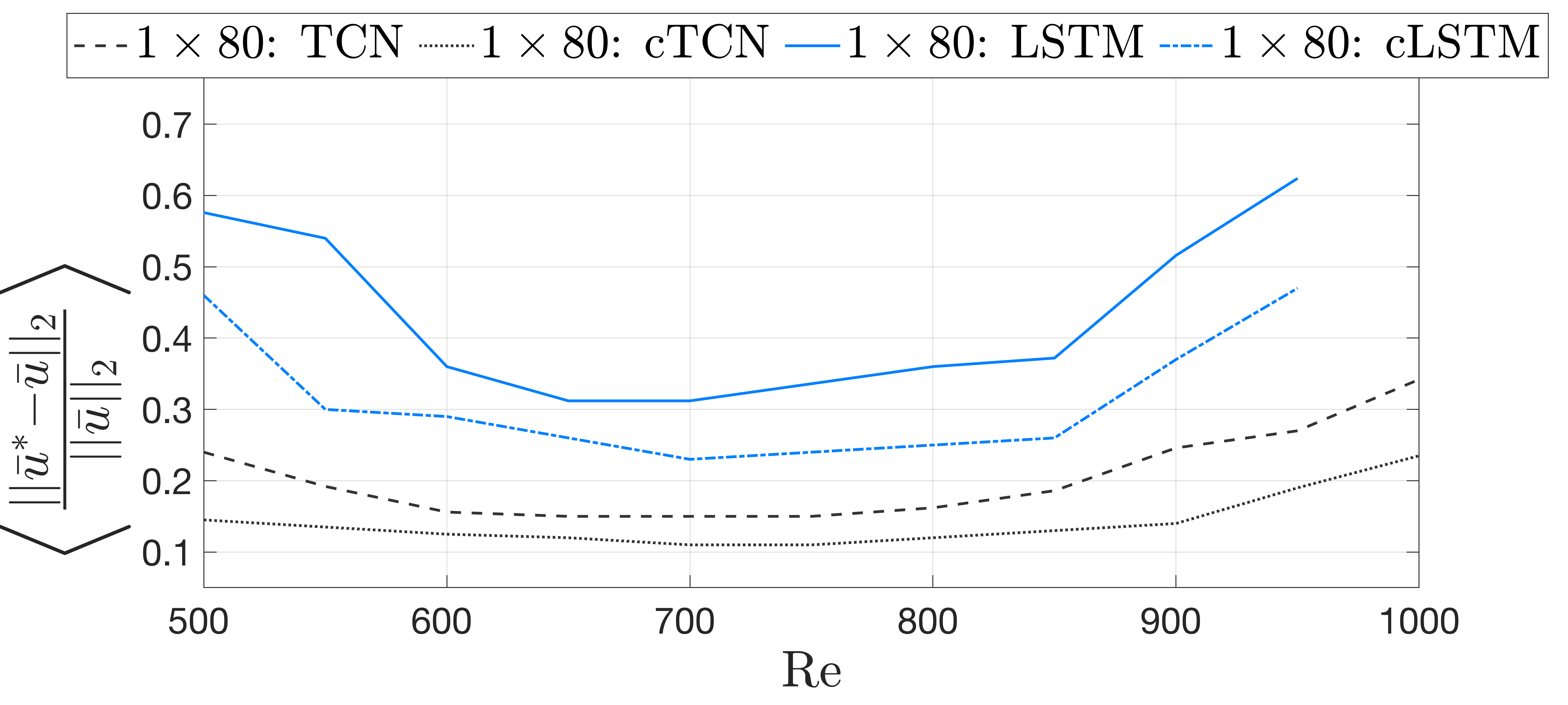}
        \caption{
        Normalized error (\ref{error1d}) for one-dimensional closure models error for one-dimensional closures applied on bimodal jets. Training data includes unimodal jets with $Re = \{650,750,850\}$.
        }
        \label{fig:L2_error_bimodal1D}
\end{figure}

Next we test the generalizability of the closure schemes presented in the previous section on bimodal jets. Once again we state that we train on unimodal jets (as described previously) while we test our scheme on bimodal jets with the unperturbed jet-structure of the fluid flow chosen as\begin{equation}
u_1 = \exp[-3 (y-0.8 \pi)^2] +\exp[-3(y-1.2\pi)^2], \ \ Re \in [500,1000].
\end{equation}In Figure \ref{fig:L2_error_bimodal1D} we present the normalized mean-square error (\ref{error1d}) between the reference solution and the one predicted by the closure model. As we can see, employing the physical constraint during training significantly improves the results.
Specifically, we note that the errors of the cTCN and cLSTM remain to the same levels observed in the unimodal jet case (i.e. we have good generalizability properties in different flows), while the unconstrained version has significantly worse performance compared with the unimodal jet case.
In addition, we observe  much better behavior of the constrained closures when we move to Reynolds numbers higher than the ones used for training. In fact, for sufficiently large Reynolds the schemes based on the unconstrained closures become unstable. 

\begin{figure}[hbt]

\center
  \subfloat[]{       \includegraphics[width=0.45\textwidth]{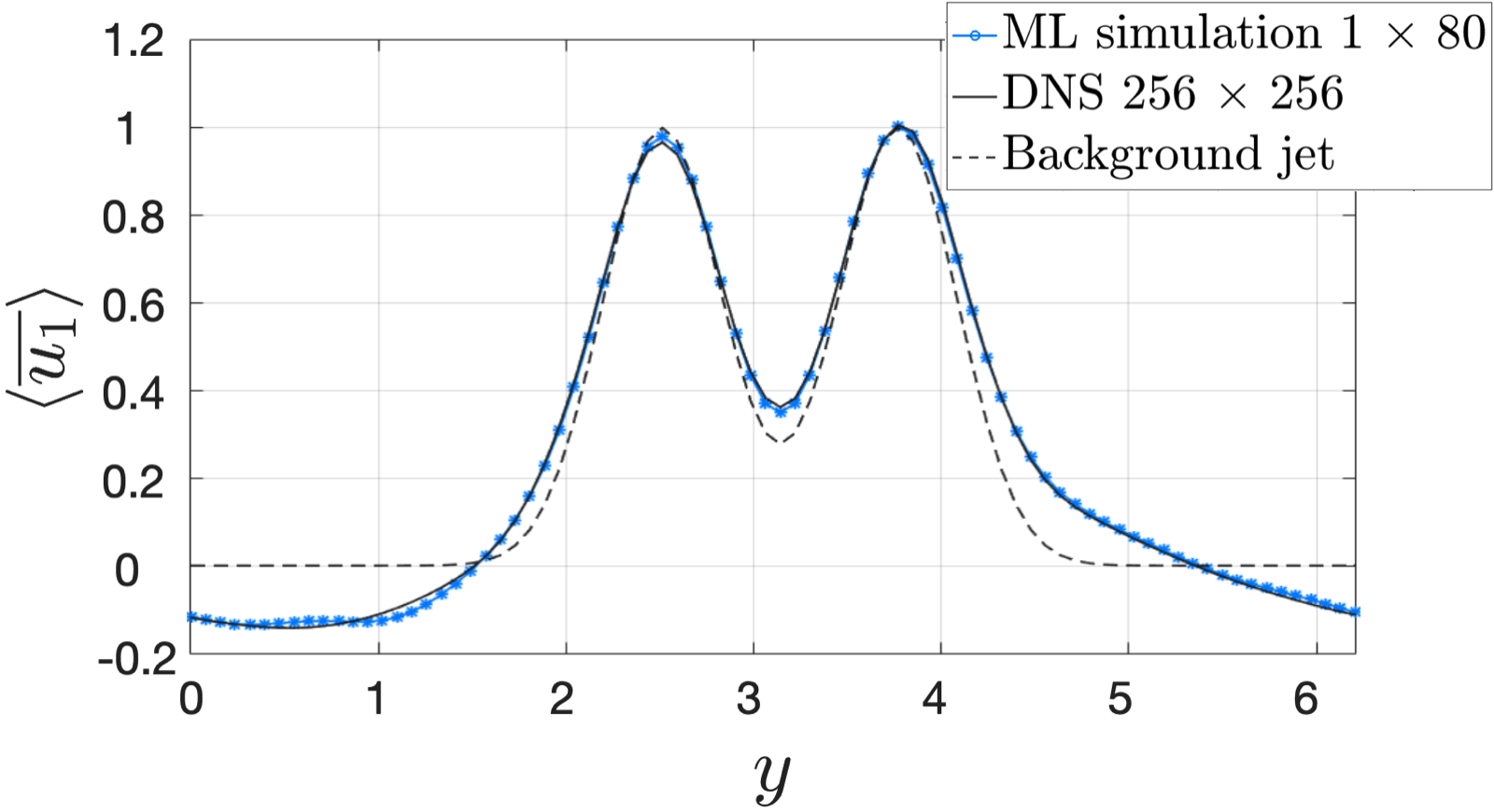}}
  \subfloat[]{ \includegraphics[width=0.45\textwidth]{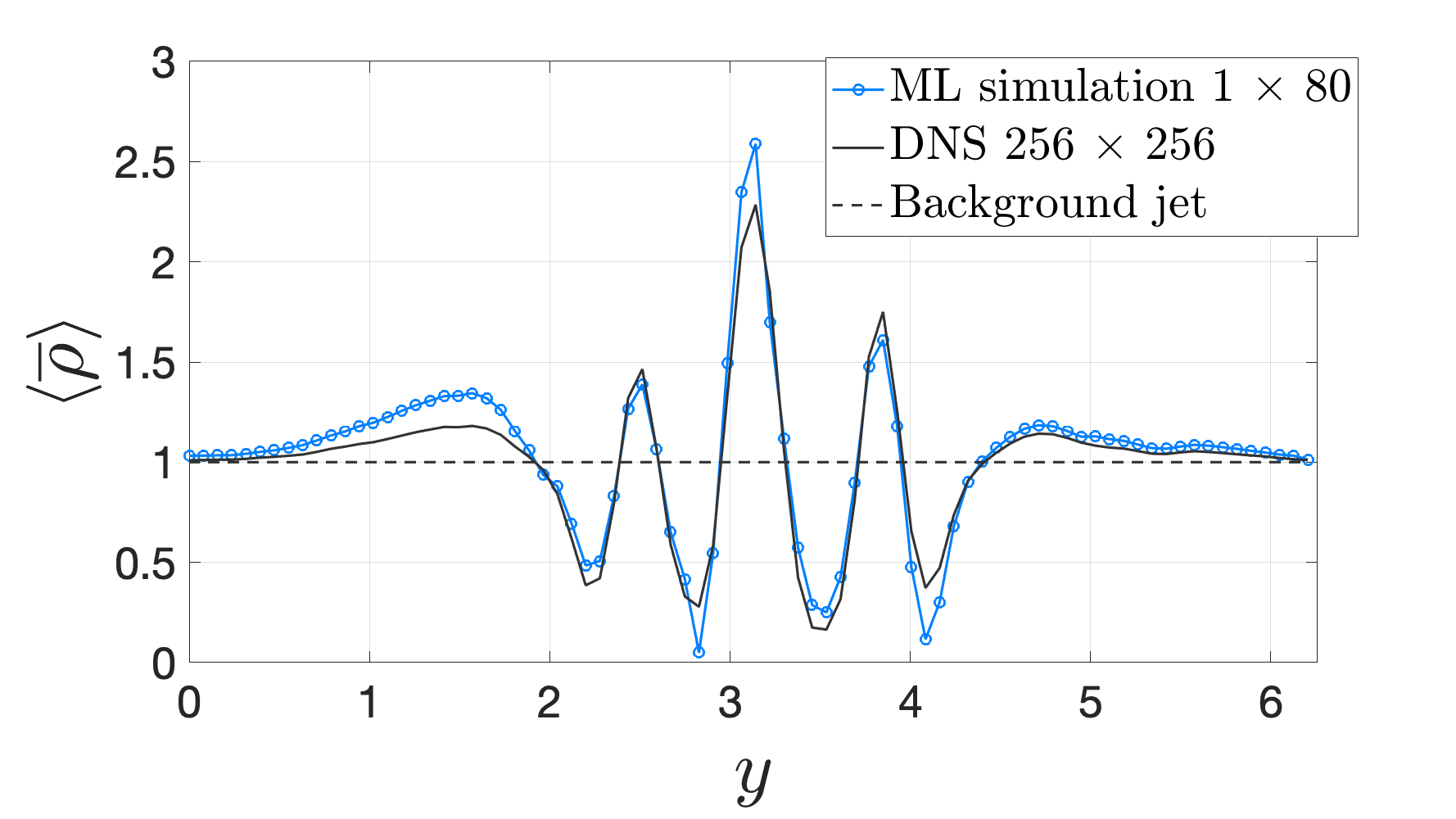}}
     \caption{
        Time-averaged profile of $\overline{u}_1$ and $\bar \rho$ for the one-dimensional cTCN
closure model (blue line) and the DNS (black line). The simulation corresponds to $Re = 1000$ and a bi-model  background jet,
while training data for the closures correspond to unimodal jets with $Re
\in
\{650,750,850\}$. The shape of the jet
that is imposed by the large-scale forcing is depicted with dashed line.
       }
        \label{Bimodal_flow1D}
\end{figure}

Results regarding the time-averaged jet profile of $\overline{u}_1$ and $\rho $ are depicted in Figure \ref{Bimodal_flow1D} for the cTCN architecture, where we can observe the excellent agreement of the predicted profile with DNS.
We also apply the closure scheme on the transport equation (\ref{Bubble_Transport})
to compute the distribution of bubbles.
We present the comparison of the mean distribution of bubbles between  the cTCN closure model and DNS in Figure \ref{Bimodal_flow1D}(b). We note that the error for this case is slightly more pronounced compared with the one observed for the mean flow velocity in Figure \ref{Bimodal_flow1D}(a). This can be attributed to two factors: i) the predictions of the transport model rely on the predictions of the coarse-scale model for the velocity field (hence error accumulates); and ii) the closure model for the bubbles relies only on data since the  energy-preserving constraint is not relevant. 

A summary of the error improvement in the one-dimensional predictions due to the adoption of the physical constraint is presented in Table 5. We present the improvement of the mean-square error for the mean flow, averaged over different Reynolds numbers. For all cases this percentage ranges between $23\%$ and $29\%$ with the improvement being more pronounced for the bimodal setups, i.e. the setup that was not used for training. 
\begin{table}[H]
\centering
\caption{Error decrease (Reynolds-averaged) due to the physical constraint for one-dimensional closure schemes. }
\begin{tabular}{|P{2cm}||P{1.5cm}|P{2.5cm}|}
 \hline
 Architecture & Jet-type & Error decrease  \\
 \hline
 TCN-1D   & Unimodal & 23\%  \\
 LSTM-1D &   Unimodal & 25\%  \\
 TCN-1D   & Bimodal    & 29\% \\
 LSTM-1D &   Bimodal   & 27\% \\
 \hline
\end{tabular}
\end{table}

\section{Validation and generalizability for two-dimensional closures}
Here we aim to showcase the application of our method on two-dimensional coarse-scale closures. As previously, we consider two cases: (i) we train on \textit{unimodal} jets for flows with Reynolds number $Re \in \{650,750,850\}$ and test on \textit{unimodal} jets in the range $Re \in [500,1000]$ and not included in the training set; (ii) we once again train on \textit{unimodal} jets (same Reynolds as before) and test on \textit{bimodal} jets in the same Reynolds range as in case (i).
For the coarse-scale model we employ a  resolution of $48 \times 48$, complemented with the ML-closure terms. We compare the energy spectra at the statistical steady state of the flows between the coarse-scale predictions and the two-dimensional reference solutions, i.e. for $t \in [200, 600]$. 
%
%

\subsection{Testing on a unimodal jet}
\begin{figure}[hbt]
\centering
        \includegraphics[width=0.75\textwidth]{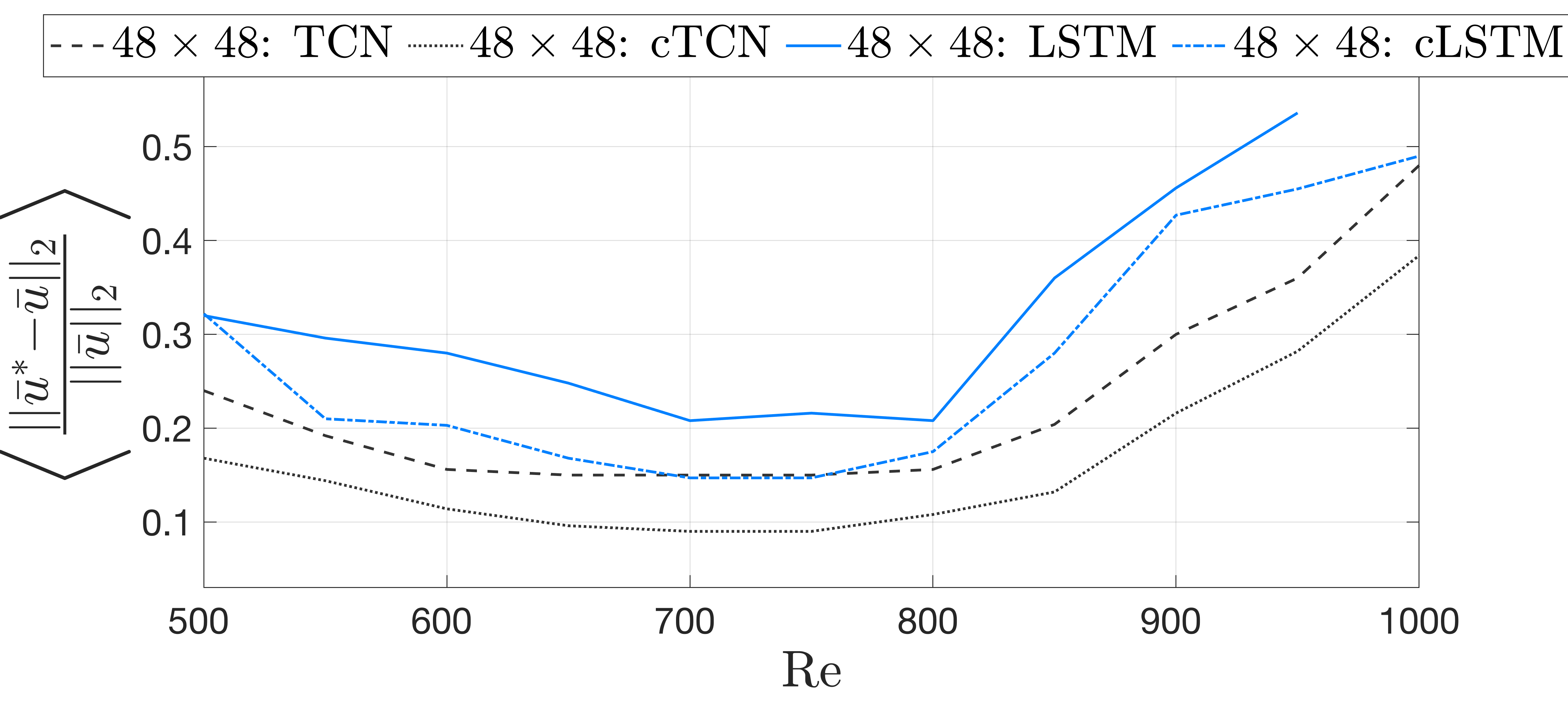}
        \caption{
        Normalized mean-square error (\ref{error2d}) of each two-dimensional closure model using TCN, LSTM and their constrained versions for unimodal jets. Training includes data for unimodal jets with $Re \in \{650,750,850\}$.
        }
        \label{fig:L2_Error_unimodal}
\end{figure}

\begin{figure}[H]
\centering
        \includegraphics[width=0.95\textwidth]{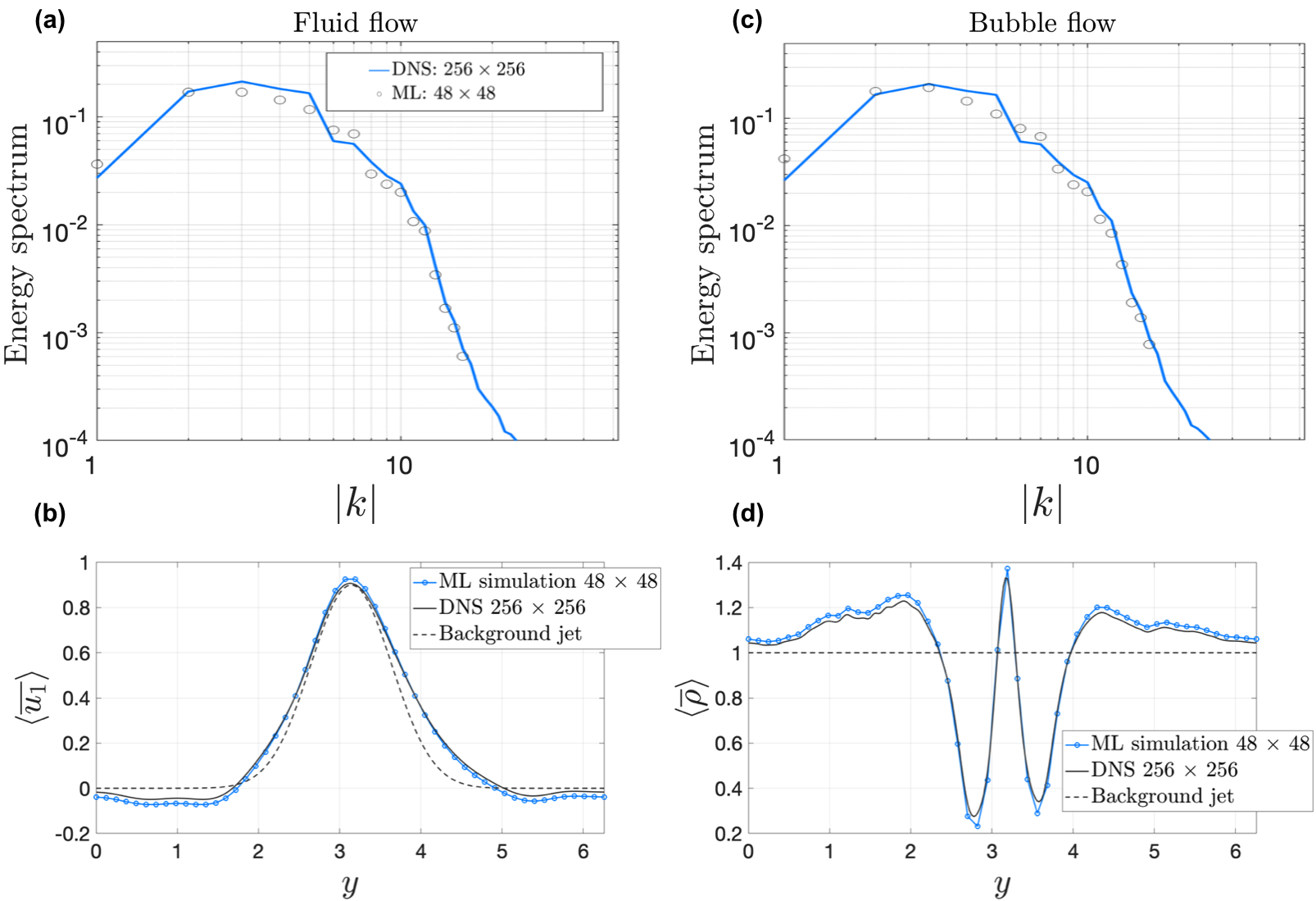}
        \caption{
        Comparison of energy spectra and mean profiles for the flow velocity field (a, c) and the bubble velocity field (b, d) for unimodal jets. Blue lines correspond to DNS simulations on a $256 \times 256$ grid and the black circles correspond to the coarse model using two-dimensional closures (cTCN) on a $48 \times 48$ grid for $Re = 800$. 
        }\label{fig:Unimodal_Energy_Spectra2D}
\end{figure}

 In Figure \ref{fig:L2_Error_unimodal} we present the space-time-averaged mean-square error between the $x-y$ locally averaged DNS flow field (using eq. (\ref{smooth2d})), $\bar u^*$, and the coarse scale model, $\bar u$:
\begin{equation}
    ||\bar u^*-\bar u||^2_2=\frac{1}{(2\pi)^2 T}\int_0^{2\pi}\int_0^{2\pi}\int_{t_0}^{t_0+T}(\bar u^*(x,y,t)-\bar u(x,y,t))^2dxdydt.
    \label{error2d}
\end{equation}
 The results are in full consistency with the one-dimensional closures, i.e. cTCN has the best performance. We also present a detailed comparison for $Re=800$ between the coarse-model and the DNS simulation in terms of the energy spectrum and mean profile the flow (Figure \ref{fig:Unimodal_Energy_Spectra2D}). The energy spectrum is computed by obtaining the spatial Fourier transform at each time instant and then considering the variance of each Fourier coefficient over time. We plot the energy spectrum  in terms of the absolute wavenumber values. For both the flow field and bubble field the coarse-model is able to accurately capture the mean profiles, as well as the large scale features of the spectrum.

\subsection{Testing generalizability on bimodal-jets}

\begin{figure}[H]
\centering
        \includegraphics[width=0.75\textwidth]{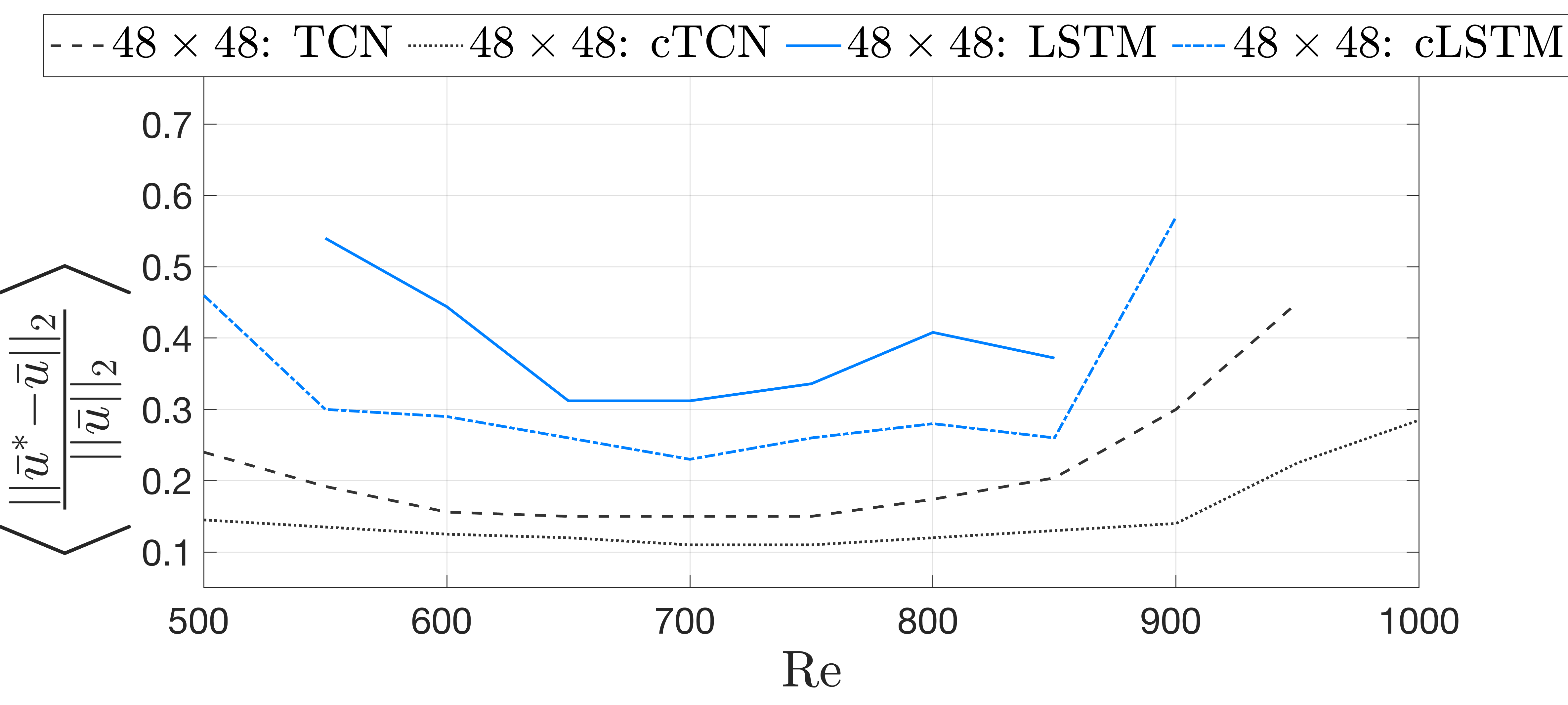}
        \caption{
        Normalized mean-square error (\ref{error2d}) for two-dimensional coarse models  applied on bimodal jet flows. Training used data from unimodal flows with $Re = \{650,750,850\}$.
        }
        \label{fig:L2_error_bimodal}
\end{figure}

We  proceed to test the generalizability of the two-dimensional closure schemes on bimodal jets. The setup is identical with the one adopted for one-dimensional closures. In Figure \ref{fig:L2_error_bimodal} we compare the noormalized mean-square error between the locally averaged DNS solution and the one obtained form the coarse model. Consistently with the previous results the cTCN has the best performance. It is interesting to note that the performance is even better in the Reynolds regime outside the training data, i.e. for $Re>850$. 

\begin{figure}[H]
        \includegraphics[width=\textwidth]{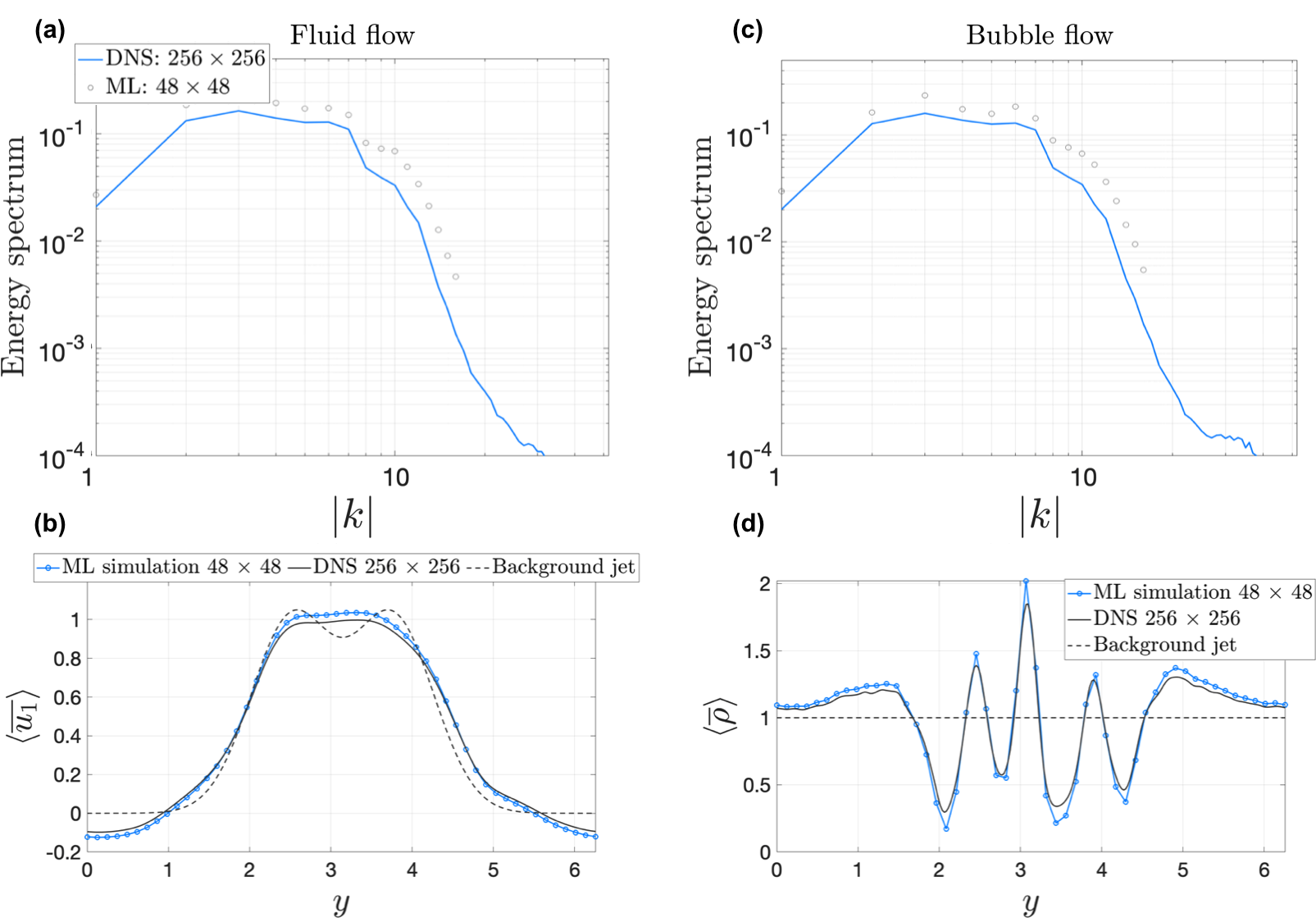}
        \caption{
        Comparison of energy spectra and mean profiles for the fluid velocity field (a, c) and the bubble velocity field (b, d) for the case of a bimodal jet with $Re = 800$. Blue lines correspond to DNS simulations on a $256 \times 256$ grid and the black circles correspond to a coarse-model with two-dimensional cTCN closures on a $48 \times 48$ grid. 
        }\label{fig:Bimodal_Energy_Spectra2D}
\end{figure}

\begin{figure}[H]
\centering
        \includegraphics[width=0.95\textwidth]{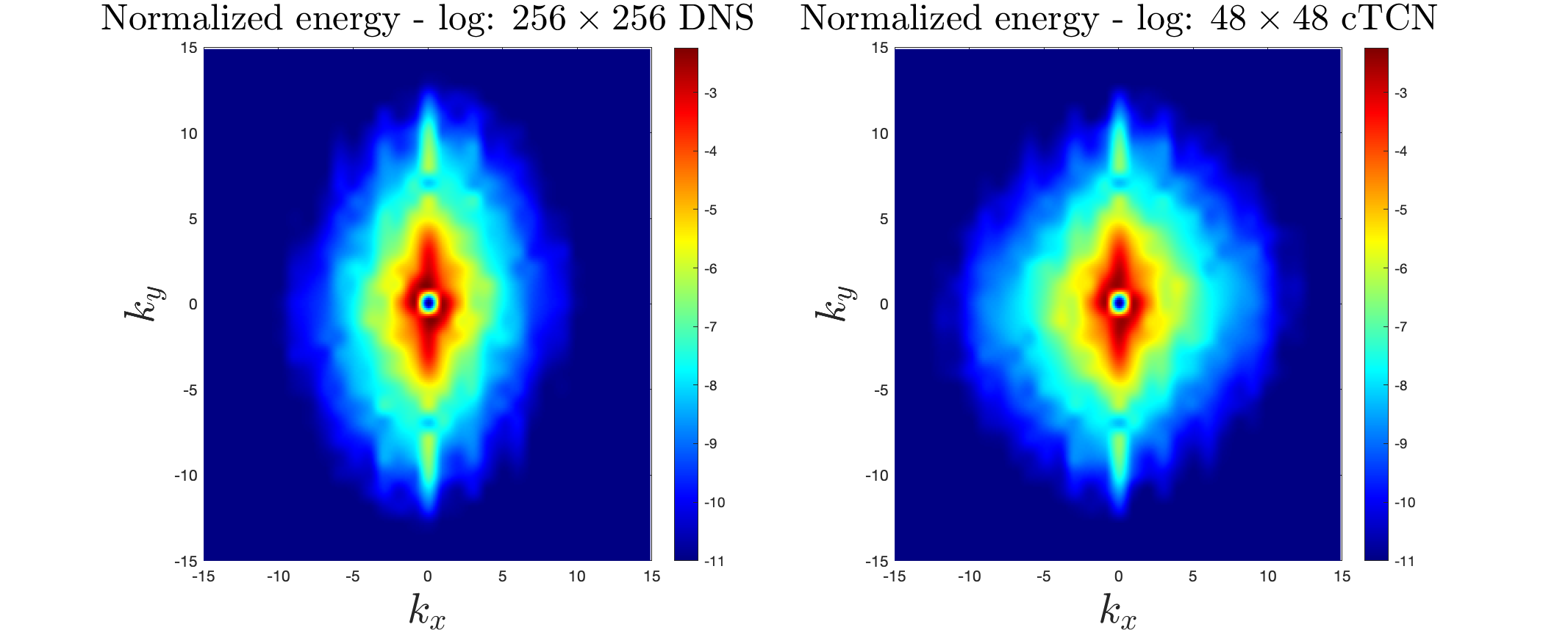}
        \caption{
        Energy spectrum of the fluid flow for the bimodal jet for $Re = 800$. Comparison between the DNS simulation (left) and the coarse-model based on 2D cTCN closures (right).
        }
        \label{fig:Energy_Bimodal}
\end{figure}

The energy spectrum of the fluid velocity and the bubble velocity field, as well as the corresponding mean profiles are compared with DNS for $Re=800$ in Figure \ref{fig:Bimodal_Energy_Spectra2D}. In this case, we note that while there is good agreement between the mean profiles, there is some discrepancy between the approximate and exact spectra. To understand better the source of this discrepancy we plot the energy spectrum of the flow in the $k_x, k_y$ space (Figure \ref{fig:Energy_Bimodal}). As it can be seen the coarse model overestimates the spread of the energy of the fluctuations only in the $x-$direction, which is consistent with the fact that the mean $y-$profile of the flow is accurately modeled. This is not surprising given that the developed closures in this paper are designed to capture well the mean flow characteristics and not necessarily the energy spectrum. A closure approach based on second-order statistical equations (see e.g. \cite{sapsis_majda_mqg}) is beyond the scope of this work and will be considered elsewhere.

The overall improvement in the two-dimensional predictions due to the adoption of the physical constraint is summarized in Table 6, where we show the improvement of the mean-square error for the mean flow, averaged over different Reynolds numbers. We note that for the TCN architecture the improvement is more pronounced, close to $30\%$, and it is also robust for the case of a bimodal jet. 
\begin{table}[H]
\centering
\caption{Error decrease (Reynolds-averaged) due to the physical constraint for two-dimensional closure schemes.}
\begin{tabular}{|P{2cm}||P{1.5cm}|P{2.5cm}|}
 \hline
 Architecture & Jet-type & Error decrease  \\
 \hline
 TCN-2D   & Unimodal & 30\%  \\
 LSTM-2D &   Unimodal & 20\%  \\
 TCN-2D   & Bimodal    & 33\% \\
 LSTM-2D &   Bimodal   & 31\% \\
 \hline
\end{tabular}
\end{table}
\subsection{Dependence on the coarse-model grid-size}

\begin{figure}[H]
\centering
       \subfloat[]{  \includegraphics[width=0.50\textwidth]{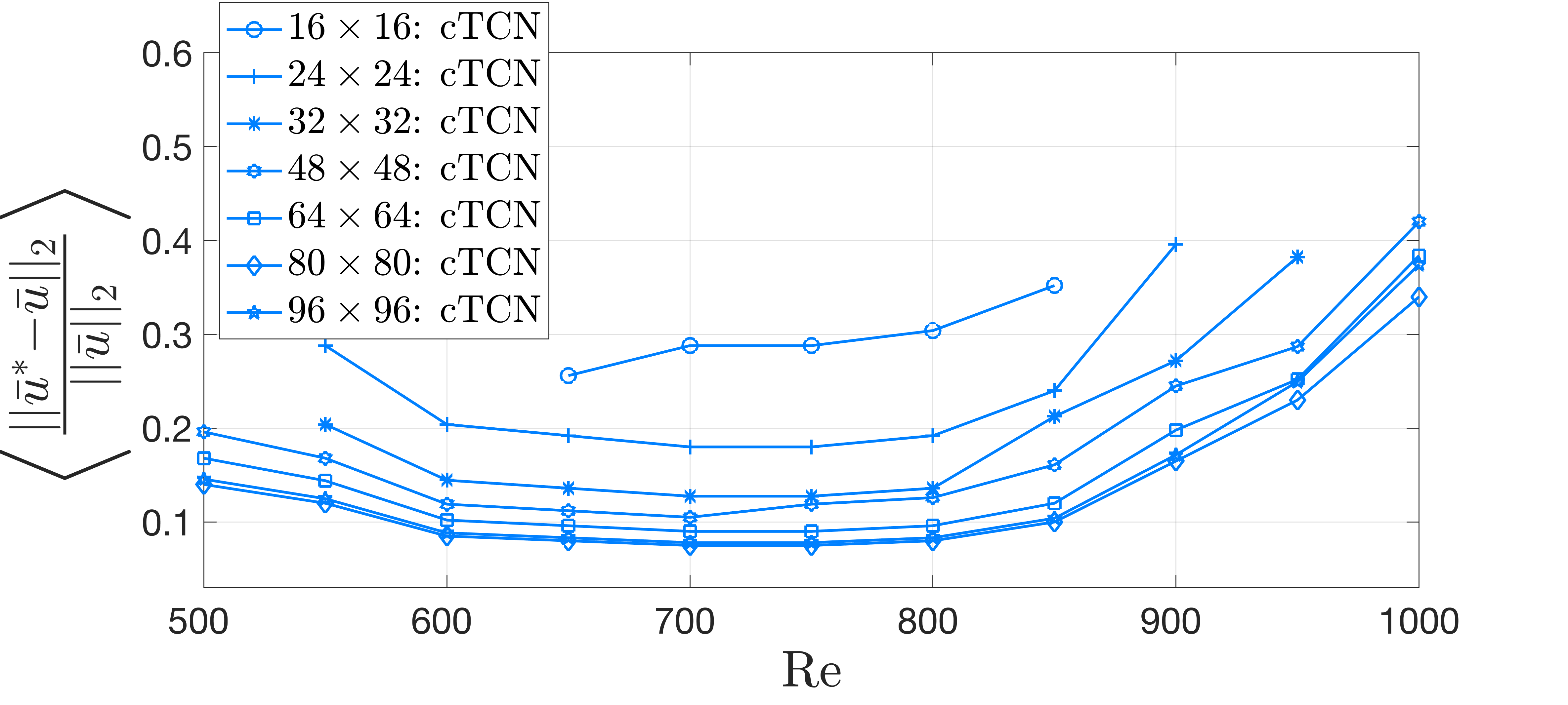}}
        \subfloat[]{  \includegraphics[width=0.50\textwidth]{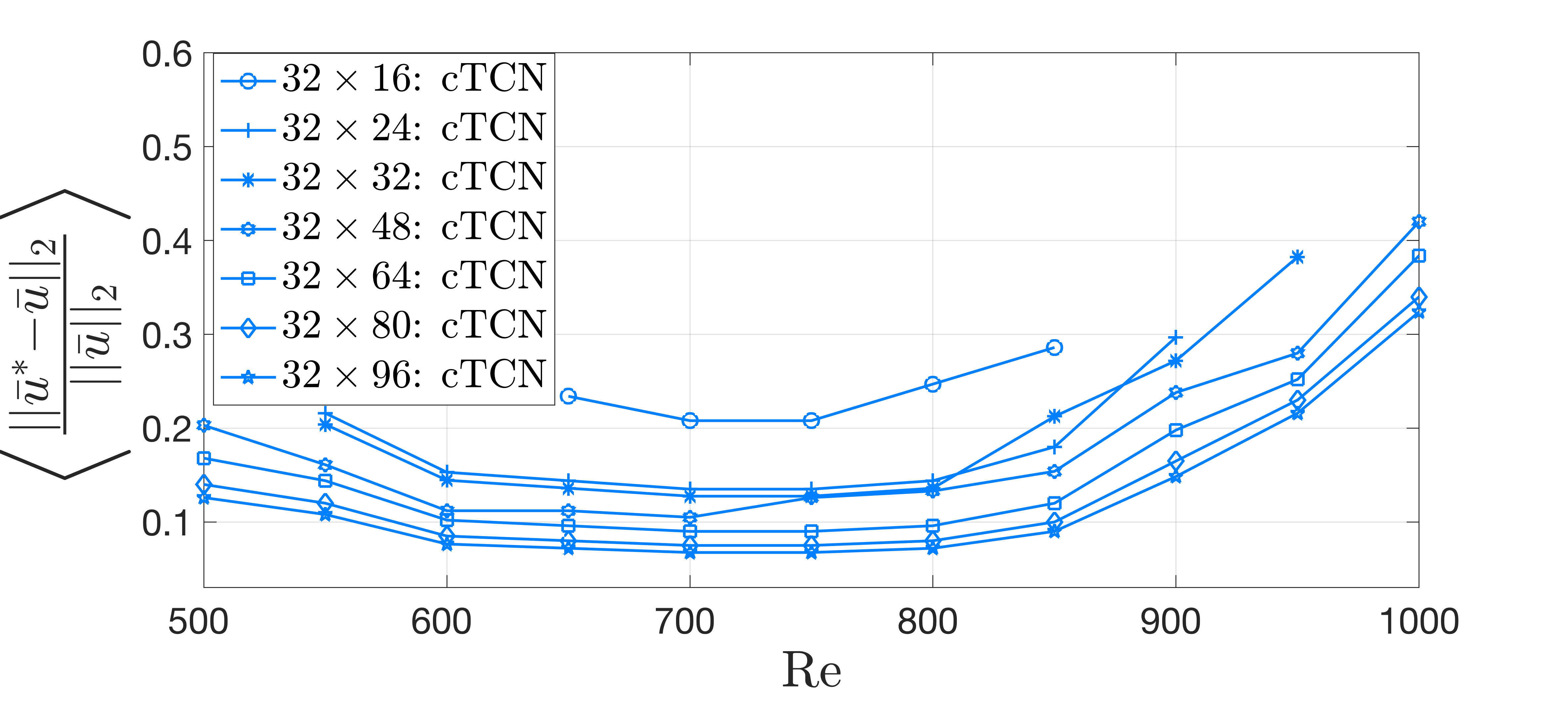}}
        \caption{
        Root-mean-square error for different grid-sizes using two-dimensional cTCN closures on bimodal jet.
        }
        \label{fig:L2_Multigrid1}
\end{figure}

Finally, we showcase a numerical investigation for the relationship between the chosen grid-size for the coarse-scale simulations and the mean-square  error of the velocity of the fluid flow. For all the results presented below, training data was chosen as previously (unimodal jets) and results are presented for bimodal jets. In Figure \ref{fig:L2_Multigrid1}(a) we vary the size of a $N_x \times N_y$ grid with $N_x=N_y \in \{16,24,32,40,48,64,80,96\}$. We notice that there is significant improvement in our predictions as we refine the grid up to a grid-size of $64 \times 64$ where the error saturates. 

Since the variation of the mean profile of the flow is only along the $y$-direction, having a coarser resolution along the less significant $x$-direction should not hinder the predictions. To validate this property we maintain a constant discretization in the $x-$ direction and vary the grid-size only in the $y-$ direction. Results are demonstrated in Figure \ref{fig:L2_Multigrid1}(b), showing clearly that by having a fine resolution only in the $y-$direction is sufficient to achieve comparable performance with the fine resolution case in both directions: $N_x=N_y=96$.

\section{Conclusions}

We have demonstrated the application of the energy conservation property of the advection terms on machine learning non-local closures for turbulent fluxes. We have adopted two neural network architectures, based on LSTM and TCN, to further include memory effects in our analysis. Clearly, the physical constraint is not restricted to these two frameworks and can be employed in other machine learning architectures, as well as other fluid systems (e.g. environmental flows). We demonstrated the computed closures in  two-dimensional jets in an unstable regime and showed that closures obtained from unimodal jets can be used for different jet geometries. The adoption of the physical constraint significantly and consistently improved the accuracy of the mean-flow predictions obtained from the corresponding coarse-scale models independently of the adopted architecture or the flow setup. This improvement was in average $26\%$ for one-dimensional closures and $29\%$ for two-dimensional closures, being notably larger for flows that were not used for training. Moreover, the constraint improved the numerical stability of the coarse-scale models especially in Reynolds numbers, which were higher than the ones included in the training data sets. While the adopted examples are relatively simple, yet unstable, fluid flows, the presented energy constraints do not depend on the complexity of the flow. They are applicable for more complicated setups including boundary flows and transition phenomena, directions that we plan to pursue in the future.

\subsubsection*{Acknowledgments} This work has been supported through the ONR-MURI grant N00014-17-1-2676 and the AFOSR-MURI Grant No. FA9550-
21-1-0058.



\bibliography{library,references} 
\bibliographystyle{abbrv}


\end{document}